\documentclass[letterpaper, 11pt]{article}

\usepackage{setspace}
\usepackage[margin=1in]{geometry}
\usepackage{array}
\usepackage{url}
\usepackage[breaklinks]{hyperref}

\hypersetup{
    colorlinks=true,
    linkcolor=blue,
    citecolor=black,
    filecolor=magenta,      
    urlcolor=blue,
}
\usepackage[normalem]{ulem}
\usepackage[hang,flushmargin]{footmisc}
\usepackage[compact,small]{titlesec}
\usepackage{microtype}
\usepackage{caption}
\usepackage{amsmath, amssymb}
\usepackage{booktabs}
\usepackage[flushleft]{threeparttablex}
\usepackage{subcaption, bm, amsmath}
\usepackage{natbib}
\usepackage{listings}
\usepackage{csquotes}
\usepackage{color, colortbl}
\definecolor{LightGray}{gray}{0.9}
\usepackage{graphicx}
\makeatletter
\newcommand\footnoteref[1]{\protected@xdef\@thefnmark{\ref{#1}}\@footnotemark}
\makeatother

\title{\Large Explaining Sustained Blockchain Decentralization with Quasi-Experiments:\\ The Resource Flexibility of Consensus Mechanisms}

\author
{
    \vspace{-0.5em}
    \normalsize Harang Ju\textsuperscript{1}, Madhav Kumar\textsuperscript{2}, Ehsan Valavi\textsuperscript{3}, Sinan Aral\textsuperscript{3}\\ \vspace{-0.5em}
    \normalsize \textsuperscript{1}Johns Hopkins Carey Business School, harang@jhu.edu\\ \vspace{-0.5em}
    \normalsize \textsuperscript{2}Harvard Business School, mkumar@hbs.edu\\ \vspace{-0.5em}
    \normalsize \textsuperscript{3}MIT Sloan School of Management, \{evalavi, sinan\}@mit.edu
}
\date{\normalsize\today}

\begin{document} 

\maketitle 

\begin{abstract}
Decentralization is a fundamental design element of the Web3 economy. Blockchains and distributed consensus mechanisms are touted as fault-tolerant, attack-resistant, and collusion-proof because they are decentralized. Recent analyses, however, find some blockchains are decentralized, others are centralized, and that there are trends towards both centralization and decentralization in the blockchain economy. Despite the importance and variability of decentralization across blockchains, we still know little about what enables or constrains blockchain decentralization. We hypothesize that the resource flexibility of consensus mechanisms is a key enabler of the sustained decentralization of blockchain networks. We test this hypothesis using three quasi-experimental shocks (policy-related, infrastructure-related, and technical) to resources used in consensus. We find strong suggestive evidence that the resource flexibility of consensus mechanisms enables sustained blockchain decentralization and discuss the implications for the design, regulation, and implementation of blockchains.
\end{abstract}

\clearpage

\section{Introduction}

Blockchains offer an alternative to centralized systems through a model of trust built on distributed consensus \citep{ammous2018bitcoin}. While they were originally conceived as a time-stamping mechanism to prevent back- or forward-dating digital documents \citep{haber1991timestamp}, in their modern incarnation, blockchain implementations are designed to remove centralized institutions from the process of tracking electronic cash \citep{nakamoto2008bitcoin} and to execute generalized software programs without intermediaries \citep{buterin2014ethereum}. Decentralization is fundamental to the ability of blockchains to operate without centralized intermediaries while maintaining resistance to attacks, faults, and collusion \citep{buterin2017meaning}. Blockchain decentralization is multidimensional, and different dimensions may benefit from different levels of centralization \citep{halaburda2020multidimensional, hsieh2023future}: while centralization can improve coordination in platform governance and system development \citep{chen2021governance}, decentralization in the consensus layer is critical for security and trust. Indeed, failure to sustain decentralization in blockchain \textit{consensus mechanisms} can cause catastrophic failures that lead to the loss of billions of dollars of economic value and data integrity regarding asset ownership. For example, the Ronin blockchain was hacked in 2022 for \$625 million because of centralized points of failure \citep{thurman2023} and 51\% attacks occur frequently and have significant and lasting negative effects on token prices \citep{shanaev2019cryptocurrency}. Since 2009, with blockchains growing rapidly and cryptocurrencies surpassing a 3.7 trillion-dollar market capitalization,\footnote{\href{https://coinmarketcap.com/charts/}{CoinMarketCap} Accessed February 18, 2025.} and with institutions, including governments, increasingly tokenizing hundreds of billions of dollars of bonds and other assets on blockchains, understanding and sustaining decentralization is becoming increasingly critical \citep{eib2021digitalbond, securitize2024}.

While many tout the decentralized nature of blockchain technology, it is still unclear whether blockchains are actually becoming more centralized or decentralized over time and, perhaps more importantly, what drives blockchain centralization or decentralization. Indeed, decentralization faces countervailing economic forces, including economies of scale and coordination costs \citep{hui2023dcai}. It also faces economic trade-offs with blockchain's other goals, such as net neutrality and permissionlessness \citep{halaburda2024neutrality, bakos2021permissioned}. Recent empirical work also estimates that several key subsystems of blockchain ecosystems are becoming more centralized \citep{ju2025decentralizing}. In short, decentralization cannot be taken for granted; it must be actively sustained against countervailing economic and technical forces. In this paper, we explore potential enablers of sustained blockchain decentralization in consensus mechanisms, a key element of this technology critical to establishing trust and fault tolerance without centralization. While prior work has identified economies of scale \citep{capponi2023pow}, governance structures \citep{chen2021governance}, and technological innovations \citep{cong2022scaling} as factors influencing decentralization, the role of the resources required for consensus participation, and specifically their flexibility, has not been empirically examined.

We hypothesize that the flexibility of resources required for consensus mechanisms is an important enabler of sustained blockchain decentralization. Resource flexibility refers to the ease with which participants can acquire, deploy, and redeploy the resources needed for consensus, from portable hardware like GPUs to transferable staked tokens (see Section~\ref{sec_theory} for a full development of this concept). 
Many blockchains require the use of costly resources to verify transactions for proof-based consensus mechanisms, including hardware and electricity for Proof-of-Work (PoW) and native blockchain tokens for Proof-of-Stake (PoS). Empirical examination of the implications of technical resource flexibility on sustained blockchain decentralization can therefore inform design decisions in the pursuit of decentralization at the heart of blockchain technologies.\footnote{The Proof-of-Authority consensus mechanism, which relies on a small number of validators selected by a central authority, is an exception to this requirement, though its inclusion in the validator set, determined by centralized authorities, is also a costly resource.}

We leverage quasi-experimental variation across six blockchains (\textit{i.e.}, Bitcoin, Ethereum, Solana, Gnosis, BNB, Ronin) to investigate how resource flexibility, proxied by differences in blockchain designs, influences blockchain decentralization in the context of three independent shocks: a policy-related shock (China’s crypto mining ban on May 15, 2021), an infrastructure-related shock (Hetzner’s shutdown of Solana validators on November 2, 2022), and a technical shock (the Ethereum Merge on September 15, 2022).
China's ban shows the impact of resource flexibility (\textit{i.e.}, ASICs vs. GPUs) in response to a policy shock; Hetzner's shutdown contrasts with China's ban by revealing the impact of an even more flexible resource (\textit{i.e.}, tokens in PoS vs. hardware in PoW) in response to an infrastructure shock; and the Ethereum Merge assesses the effect of increased resource flexibility as Ethereum upgraded from PoW to PoS. To measure these effects, we employ event studies, difference-in-difference (DiD) estimation, and synthetic DiD (SDiD) methods and use five different measures of decentralization: Shannon entropy, the number of validating nodes, the Gini coefficient, the Nakamoto coefficient, and the Herfindahl–Hirschman Index. 

Our analysis provides strong evidence suggesting that the resource flexibility of distributed consensus mechanisms enables sustained blockchain decentralization. We find that China's ban reduced Bitcoin's decentralization in comparison to Ethereum's by around 0.21 bits of Shannon entropy from around 3.7 bits, that Hetzner's shutdown reduced Solana's decentralization by around 0.38 bits and that the dynamics of the recovery from the initial shocks varied widely across the two shocks. Bitcoin's decentralization took over 6 months to recover from China's ban, while Solana's decentralization fully recovered from the Hetzner shutdown in approximately 49 days. Our analysis also decomposes the entropy into the number of nodes in the network and the distribution of block production across those nodes. We discuss why the difference in recovery rates between Bitcoin and Ethereum in response to the China ban and Solana's rapid recovery from the Hetzner shutdown together support our hypothesis. We also find that the Ethereum Merge, which transitioned Ethereum from a Proof-of-Work (PoW) consensus mechanism (which is less flexible) to a Proof-of-Stake (PoS) mechanism (which is more flexible), provides a natural proxy for increased resource flexibility, thereby increasing Ethereum's decentralization by 1.27 bits. Collectively, our findings offer strong suggestive evidence that variation in the level of resource flexibility in consensus mechanisms, from ASICs to GPUs and from PoW to PoS, is associated with sustained levels of blockchain decentralization.

To the best of our knowledge, ours is the first study to empirically examine the role of consensus mechanism design in sustaining blockchain decentralization and our work contributes to several areas of research. First, our study is related to the literature on blockchain consensus mechanisms and their economic and technical implications \citep{bamakan2020survey, halaburda2022microeconomics}. Consensus predicates many of the unique features of blockchains, including permissionlessness and trustlessness, which in turn support ownership, native payments, DeFi, and arbitrary smart contracts without central authorities \citep{catalini2016simple, cong2019blockchain, harvey2021defi, john2022bitcoin}. While previous studies illustrated the benefits of the technology, our paper is among the first to empirically demonstrate whether and how the consensus mechanism supports blockchain technology's promise of a decentralized network.

Our work also contributes to the literature on the decentralization of blockchains \citep{gencer2018decentralization, hoffman2020towards, wu2020coefficient, cong2021decentralized, sai2021taxonomy, jia2022measuring}. Many studies have pointed out the centralized nature of blockchain consensus mechanisms \citep{srinivasan_lee_2017}, especially in permissionless blockchains \citep{bakos2021permissioned}. \cite{cong2022scaling} recently identified off-chain computation (in Layer 2 ``roll-ups") as a driver of decentralization in data providers for oracles, though not for the central mechanism of blockchain consensus. \cite{capponi2023pow} theoretically demonstrate how hardware-based consensus mechanisms can support decentralization. Interestingly, our work complements that of \cite{garratt2023fixed} who show that fixed costs in crypto mining increase resilience against 51\% attacks when there are shocks to the price of crypto mining rewards. They argue fixed costs make it harder for miners to switch their hardware, or other resources, to other uses when mining rewards drop in price, thus improving resilience against attacks. Our results show the flip side of resource flexibility: when there are shocks \textit{to the resources} required for mining (\textit{versus} \textit{the rewards} from mining), it is harder for miners or validators to obtain or migrate the resources to keep mining. Given the \textit{raison d'\^{e}tre} of blockchains of forgoing centralized intermediaries and resisting external shocks, it is essential to understand the conditions under which blockchains are or are not decentralized and shock resistant, especially as more economic activity occurs on blockchains, including trading in government bonds \citep{eib2021digitalbond} and stock exchanges \citep{meichler2023}. We believe our work reveals the nuanced role of resource flexibility in sustaining decentralization in the consensus mechanisms at the heart of blockchains. Table~\ref{tb_literature} in the Appendix provides an overview of key studies on blockchain decentralization and positions our contribution within this literature.

\section{Theory: Decentralization and Resource Flexibility} \label{sec_theory}

\subsection{Decentralization}

The underlying integrity of blockchains depends on their decentralization, which ensures transaction validity without trusted intermediaries by preventing faults, attacks, and collusion \citep{buterin2017meaning}. We define decentralization as the distribution of decision-making authority and control across independent entities \citep{vergne2020decentralized, garratt2023fixed, cong2021decentralized}. While higher participation can promote greater decentralization \citep{mueller2024understanding}, decentralization is ultimately a question of how power and control are distributed. Thus, we use multiple measures of decentralization to capture meaningful shifts in the consensus layer, not participation alone. Despite its importance, empirical evidence shows significant variation in decentralization across blockchains and over time \citep{gencer2018decentralization, sai2021taxonomy}, underscoring the need to examine the factors driving these variations.

Importantly, blockchain decentralization is not monolithic. \cite{halaburda2020multidimensional} argue that blockchain governance involves multiple dimensions, each of which may benefit from different levels of centralization. While centralization can improve coordination in system development \citep{chen2021governance}, decentralization in transaction validation is essential for the security guarantees that blockchains promise. Similarly, \cite{hsieh2023future} distinguish among dimensions of decentralization and find that they differently affect early-stage platform growth. Our study focuses on the consensus layer, the mechanism by which blockchains validate transactions and maintain ledger integrity, where sustained decentralization is most critical for preventing attacks, faults, and collusion.

Several factors have been identified as drivers of consensus-layer decentralization. Economies of scale in mining and validation can concentrate resources among large operators \citep{capponi2023pow, cong2021decentralized}. Token distribution and wealth concentration shape validator power in Proof-of-Stake systems \citep{mueller2024understanding}. Technological innovations such as Layer-2 scaling can reduce participation frictions \citep{cong2022scaling}. Fixed costs in mining can paradoxically improve resilience by committing miners to the network when rewards decline \citep{garratt2023fixed}. Yet one factor that cuts across consensus designs remains largely unexplored empirically: the flexibility of the resources that consensus mechanisms require. Unlike governance rules or token distributions, which vary within a given consensus design, resource flexibility is determined by the consensus mechanism itself: the fundamental design choice that dictates whether participants invest in application-specific hardware, general-purpose hardware, or transferable digital tokens. This design-level property determines the switching costs participants face when disrupted, the barriers new entrants must overcome, and the speed with which the network can reconstitute after a shock. Understanding resource flexibility is therefore critical not only for explaining observed variation in decentralization, but also for informing the design of future consensus mechanisms.

At the theoretical limit, full decentralization is impossible in a permissionless system without an identity management system that gives one vote to each person \citep{kwon2019impossibility, bakos2021permissioned}. However, even in permissionless systems, \cite{cong2019blockchain} and \cite{cong2022scaling} show that technological innovations that reduce frictions inherent in participation can increase competition in mining and decentralization in automated ``oracles" that enable blockchains to seek and append information from off-chain sources. A key factor determining how effectively these frictions can be overcome is the flexibility of the resources required for consensus participation.

\subsection{Resource Flexibility}

We define resource flexibility as the ease with which participants can acquire, deploy, and redeploy the resources required for consensus across different locations and operational contexts. This flexibility is distinct from resource efficiency: while some blockchains may require higher computational resources, their ability to reallocate these resources in response to shocks is what determines their flexibility, not their absolute efficiency. To rank different resources along a flexibility gradient, we draw on the asset specificity dimension of \cite{williamson1985} transaction cost framework, which measures the switching costs of redeploying an asset to an alternative use. Resource flexibility corresponds inversely to asset specificity. Highly specific assets, such as ASICs designed solely for a single blockchain's hash function, entail high switching costs and cannot be redeployed when disrupted. General-purpose hardware like GPUs and digital assets like staked tokens can be reallocated at progressively lower cost. This gradient of asset specificity structures our empirical comparisons: from ASICs (high specificity, high switching costs) to GPUs (moderate specificity) to staked tokens (low specificity, low switching costs). Whether this flexibility translates into sustained decentralization is an empirical question, one that we test using quasi-experimental variation across blockchains with different levels of resource flexibility.

Resource flexibility can influence blockchain decentralization through two mechanisms. First, flexible resources facilitate rapid redeployment in response to disruptions: when consensus participants are forced to relocate or restructure their operations, the switching costs of their resources determine how quickly the network can recover, thereby promoting sustained decentralization. Second, flexible resources lower barriers to entry by reducing the cost and expertise required to participate in consensus, broadening the base of participants. In permissionless blockchain networks, where it is essential to prevent control concentration and mitigate centralization risks \citep{hui2023dcai}, lower barriers can promote a more robust and decentralized ecosystem.

To see how these mechanisms operate in practice, consider how individual consensus participants respond when disrupted. When an exogenous shock forces participants to adapt, whether through regulatory action, infrastructure loss, or a protocol upgrade, the resources they control determine which options are feasible and at what cost. ASIC miners, whose hardware is designed for a single hash function, face high switching costs: they cannot repurpose their equipment for alternative uses, must physically relocate heavy machinery to new jurisdictions, and face extended downtime during any transition. Only the largest, best-capitalized operators can absorb these costs, which concentrates the network among survivors and slows recovery. GPU miners face lower switching costs: their general-purpose hardware can be resold, repurposed for other computations, or migrated to cloud infrastructure, enabling a broader set of participants to resume operations. PoS validators holding staked tokens face the lowest friction: because the consensus resource is digital and instantly transferable, validators can redelegate to alternative infrastructure providers within hours. These participant-level incentives determine whether a disruption causes lasting centralization, where only the largest operators survive, or a temporary dip followed by broad-based recovery.

The relationship between resource flexibility and sustained decentralization, however, is theoretically ambiguous and cannot be assumed \textit{a priori}. While the arguments above suggest flexibility should support decentralization, there are compelling counterarguments. First, resource flexibility could enable well-capitalized actors to rapidly scale their operations and dominate the network, thereby \textit{reducing} decentralization. If flexible resources can be quickly acquired and deployed, large players with capital advantages may consolidate control faster than in systems with inflexible resources that constrain rapid expansion. Second, highly flexible resources might attract transient participants who enter during favorable conditions but exit quickly when conditions change, creating instability rather than sustained decentralization. Third, lowering barriers to entry through flexibility does not guarantee a more even distribution of power; it may simply allow more participants while concentration among top validators remains unchanged. These counterarguments underscore why the effect of resource flexibility on decentralization is an empirical question that cannot be resolved through theory alone.

Given these competing forces, we hypothesize that blockchains whose consensus resources have lower asset specificity will sustain higher levels of decentralization when subjected to exogenous shocks. This hypothesis is testable through our three quasi-experimental settings, and importantly, it is falsifiable: if resource flexibility does not matter, we should observe no differential recovery between GPU-based Ethereum and ASIC-based Bitcoin after the China ban, no rapid recovery for token-based Solana after the Hetzner shutdown, and no increase in decentralization after the Ethereum Merge. The resource flexibility framework yields specific predictions for each setting. For the China ban, we predict that Ethereum (GPU-based, more flexible) should recover decentralization faster than Bitcoin (ASIC-based, less flexible) because GPUs can be more easily relocated or redeployed to cloud infrastructure. For the Hetzner shutdown, we predict rapid recovery for Solana because staked tokens can be instantly transferred to validators on alternative infrastructure. For the Ethereum Merge, we predict increased decentralization because the transition to PoS lowers barriers to entry by eliminating hardware requirements. We test these predictions empirically and return to them when interpreting our results.

\section{Methodology} \label{sec_methodology}

\subsection{Experimental Setting} \label{sec_experimental_setting}

We examine three events that represent different dimensions of how resource flexibility influences blockchain decentralization. While the China ban provides a test of mining flexibility under a policy shock, the Hetzner shutdown and Ethereum Merge allow us to explore how infrastructure constraints and technical upgrades, respectively, affect decentralization through varying degrees of resource flexibility. Table~\ref{tb_setting} outlines the three events, the types of shocks, and the resources used in the consensus mechanisms in the events.

Because resource flexibility cannot be directly measured, we leverage these quasi-experimental shocks to reveal differential impacts across systems with varying levels of flexibility. Specifically: (1) the China ban compares ASIC-based Bitcoin (inflexible, single-purpose hardware) to GPU-based Ethereum (flexible, general-purpose hardware) under the same policy shock; (2) the Hetzner shutdown tests token-based Solana (highly flexible, instantly transferable) under an infrastructure shock, which we compare to hardware-based recovery dynamics; and (3) the Ethereum Merge observes a single blockchain transitioning from less flexible (PoW hardware) to more flexible (PoS tokens) resources. This design allows us to infer the role of resource flexibility from observed differences in decentralization outcomes.

\begin{table}[ht]
\centering
\resizebox{\textwidth}{!}{
\begin{threeparttable}
\begin{tabular}{cccccc}
\hline \\[-1.8ex]
\textbf{Event} & \textbf{Type of Shock} & \textbf{Date} & \textbf{Flexibility} & \textbf{Switching Cost} & \textbf{Method} \\
\hline \\[-1.8ex]
China bans crypto & Policy & May 15, 2021 & Low (ASIC) vs. & ASICs: single-use, physical & Diff-in-diff \\
 mining & & & Moderate (GPU) & relocation; GPUs: repurposable & \vspace{-1em} \\\\\
Hetzner shuts down & Infrastructure & Nov 2, 2022 & High & Instant digital & Synthetic diff-in-diff\\
Solana nodes & & & & redelegation & \vspace{-1em} \\\\
Ethereum Merge & Technical & Sep 15, 2022 & Moderate & Eliminates hardware & Event study\\
& & & $\to$ High & dependency & \vspace{-1em} \\\\
\hline
\end{tabular}
\begin{tablenotes}
\small
\item Notes: Flexibility reflects inverse asset specificity (Section~\ref{sec_theory}). Low = application-specific hardware (ASICs); Moderate = general-purpose hardware (GPUs); High = transferable digital tokens.
\end{tablenotes}
\end{threeparttable}
}
\vspace{0.4 em} \caption{Experimental setting and resource flexibility} \label{tb_setting}
\end{table}

\paragraph{China's ban on crypto mining.}
First, we study the policy shock of China's ban on crypto mining on May 15, 2021, analyzed in Section~\ref{sec_results_china} \citep{shen2021china}. This official Chinese policy established a blanket ban on all mining activities, including Bitcoin and Ethereum, both of which used PoW validation at the time. Both blockchains were relatively mature when the ban was enacted, with Bitcoin online since January 2009 (BTC market cap of over \$809 billion) and Ethereum online since July 2015 (ETH market cap of \$349 billion).\footnote{See \href{https://www.coingecko.com/}{coingecko.com} Accessed August 21, 2023.} This policy was a rolling ban starting around May 15, 2021, and enforced by local authorities in the next months \citep{shen2021china}. While the exact dates of the enforcement are not public, the hashrates dropped to a trough at the start of July 2021 (Figure~\ref{fig_hashrate}).

While the ban applied to both Bitcoin and Ethereum, the key difference between the chains was the hardware used to achieve consensus, which was directly related to their resource flexibility and, in turn, affected their recovery from the shock. Apart from the hardware requirements, Bitcoin and Ethereum both used PoW for consensus. We use this quasi-experimental variation in the shock to mining for Bitcoin and Ethereum to identify the role of resource flexibility in response to a large policy shock on blockchain decentralization. The technical specification underlying the variation was that Bitcoin used (and still uses) the SHA-256 hash function \citep{nakamoto2008bitcoin} and Ethereum used Ethash.\footnote{\label{fn_ethash}See a detailed explanation of Ethash on \href{https://ethereum.org/en/developers/docs/consensus-mechanisms/pow/mining-algorithms/ethash/}{ethereum.org}. Accessed October 13, 2023.} Bitcoin's SHA-256 function was primarily mined using application-specific integrated circuits (ASICs), while Ethereum's Ethash was designed to be ASIC-resistant and thus was mined primarily using graphical processing units (GPUs).\footnoteref{fn_ethash} Because ASICs are highly specialized and physically constrained to a single function, they limit adaptability in response to disruptions. In contrast, GPUs are more general-purpose, allowing for greater flexibility; miners could reallocate their computational power dynamically, including leveraging cloud-based resources. Given this variation in the inherent resource flexibility of the technologies supporting the consensus mechanisms underlying Bitcoin and Ethereum, the China ban provides an excellent natural experiment with which to test the role of resource flexibility in blockchain decentralization. For this shock, we analyze data from July 19, 2020, to March 11, 2022.\footnote{Note that the Bitcoin halving occurred on May 11, 2020, which does not confound our analysis.} 

While Bitcoin and Ethereum differ in their applications, the consensus layer is agnostic to the application layer; thus, the economic incentive structure is equivalent for both blockchains apart from the different hash functions and required mining hardware. We test the robustness of our results with consensus layer covariates (e.g., hashrates, transaction fees, token prices) in Section~\ref{sec_appendix_consensus_covar} of the Appendix and find that our results are robust to these potential confounders.

\paragraph{Hetzner's shutdown of Solana nodes.}
Second, we study Hetzner's shutdown of Solana nodes on November 2, 2022, as described in Section~\ref{sec_results_hetzner} \citep{avan_nomayo2022solana}. Hetzner is a cloud service provider based in Gunzenhausen, Germany. Operators of PoS validators often run their validators on cloud services because of their ease of deployment, high uptime, and low maintenance. Hetzner, a German cloud service provider, had explicitly prohibited any cryptocurrency-related activities on its platform as of August 23, 2022.\footnote{\label{fn_hetzner_statement} Hetzner's official statement: ``Using our products for any application related to mining, even remotely related, is not permitted. This includes Ethereum. It includes proof of stake and proof of work and related applications. It includes trading." See their \href{https://www.reddit.com/r/hetzner/comments/wucxs4/comment/ilfoj8u/}{official statement} and \href{https://x.com/Hetzner\_Online/status/1563083561975984130}{X post}. Accessed April 14, 2024.} Despite this general ban, Hetzner specifically targeted and shut down Solana nodes, while not applying the same measure to other blockchains like Ethereum. On the day of the shutdown, the delinquency rate among Solana validators spiked to 19.8\%, indicating that validators representing nearly a fifth of actively staked SOL could not validate.\footnote{\href{https://grafana.rockawayx.com/d/U8VH9rhGk/rockawayx-solana-validator?orgId=1&viewPanel=51&from=1664596800000&to=1669870799000}{RockawayX Solana Dashboard}. Accessed August 22, 2023.} Hetzner’s shutdown of Solana nodes serves as a test of how infrastructure-level constraints interact with resource flexibility in a proof-of-stake system. Unlike the China ban, which targeted mining hardware, this event highlights how validator decentralization is affected when centralized service providers restrict access, underscoring the role of infrastructure flexibility in blockchain resilience. We leverage synthetic difference-in-difference, using a synthetic control of other PoS blockchains, and event study methods to aid relevant comparisons and support identification. For this shock, we analyzed data from September 13, 2022, to December 22, 2022, 50 days before and after the shock on November 2, 2022.

This selective enforcement by an infrastructure provider represents an exogenous shock to the Solana network, offering a valuable perspective on infrastructure dependency and the risks posed by centralized cloud services to blockchain decentralization (see Section~\ref{sec_appendix_hetzner_details} for discussion of why Hetzner may have targeted Solana specifically and Figure~\ref{fig_sol_delinquency} for the delinquent stake ratio during the shutdown).

\paragraph{The Ethereum Merge.}
Third, we study the Ethereum Merge on September 15, 2022, as described in Section~\ref{sec_results_merge} \citep{EthereumMerge2023}. The Paris Upgrade, as it is also known, merged Ethereum's original blockchain with the new PoS consensus mechanism \citep{paris2022}. At that point, each new block was no longer mined by PoW miners but was from then on validated by PoS validators. While in PoW, miners used servers with graphical processing units (GPUs) to mine new blocks through energy-intensive computations, anyone after the PoS upgrade could start validating on any server using 32 ETH. This upgrade is perhaps the biggest technical shock to Ethereum's consensus mechanism to date. We analyze this shock as an event study to examine the effect of a major software upgrade from PoW to PoS on blockchain decentralization. For this shock, we analyze data from January 1, 2022, to April 1, 2023.

The Merge can be treated as an exogenous technical shock: it occurred in under 20 seconds, had been delayed five times since 2017, and was of unprecedented scale in the blockchain ecosystem, making its effects difficult to anticipate (see Section~\ref{sec_appendix_merge_exogeneity} for detailed justification).

\subsection{Data} \label{sec_data}

We used node-level data from six different blockchains to study the drivers of blockchain decentralization. A blockchain is a linked list of blocks, each containing a list of transactions. A node is a server that operates as a validator in PoS consensus or a miner in PoW consensus. In general, around every $N$ seconds, depending on the blockchain specifications, a node produces a block that is appended to the blockchain.

The data we use consists of the daily number of blocks produced by each node. Here, we focus on six blockchains (Bitcoin, Ethereum, Solana, BNB, Ronin, and Gnosis) that support quasi-experimental identification. Bitcoin, Ethereum, Solana, and Gnosis are all \textit{permissionless} blockchains in the sense that anyone can operate a node and join the network to produce blocks on the blockchains. However, the four blockchains vary in their designs, which allowed us to tease apart potential drivers of decentralization. Bitcoin is still PoW and uses specialized hardware (i.e., application-specific integrated circuits \textit{ASICs}) and energy-intensive computations to achieve consensus. Ethereum relied on PoW until September 15, 2022, when it was upgraded to PoS, which uses staked native tokens (\textit{i.e.}, ETH) to achieve consensus. Solana also uses PoS along with Proof-of-History (PoH), which scales transaction throughput while maintaining transaction sequences.\footnote{While important, we do not take into account potential Sybil attacks in our analysis, as no scalable anti-Sybil systems exist for blockchains or are in development \citep{mohan2024sybil}. See Vitalik Buterin's blogs on \href{https://vitalik.eth.limo/general/2019/11/22/progress.html}{hard problems in cryptocurrency} and \href{https://vitalik.eth.limo/general/2023/07/24/biometric.html}{proof of personhood}. Hence, our measures of decentralization are technically the best-case approximations of decentralization. See Section~\ref{sec_limitations} for further discussion on Sybil concerns.} Gnosis, an Ethereum sidechain originally designed for fast and low-cost transactions, previously used Proof of Authority (PoA) with a limited set of validators but transitioned to PoS on December 8, 2022, which uses staked native tokens (\textit{i.e.}, GNO) to achieve consensus.

In contrast to the other chains, Ronin and BNB are \textit{permissioned} blockchains that have both centrally designated and elected validators.\footnote{Ronin used Proof of Authority (PoA), which has a limited number of validators chosen by a central entity (often a company), but transitioned to Delegated Proof of Stake (DPoS), which has a limited number of elected and centrally designated validators, in April 2023.} Importantly, permissioned blockchains can be affected by policy-related or infrastructure-related shocks. Hence, we use the permissioned PoS blockchains as control units for Solana in the analysis of Hetzner's shutdown (an infrastructure shock) and not for the technical shock of the Ethereum Merge, where permissioned blockchains may be less relevant.

The data for Bitcoin and Ethereum hashrates were obtained from \href{https://blockchain.com}{blockchain.com} and \href{https://etherscan.io}{etherscan.io}, respectively (see Section~\ref{sec_appendix_data_details} for details on hashrate measurement). We aggregate daily counts of miners per day for Bitcoin, Ethereum, BNB, Gnosis, and Solana on Dune Analytics (\href{https://dune.com/}{dune.com}). Table~\ref{tb_data} shows a summary of the blockchains and data used in our study. We obtained the same data for Ronin from Moku (\href{https://moku.gg}{moku.gg}). For Bitcoin, we divided the attribution of the block proportionally among the recipients of the block reward to account for mining pools that distribute rewards in the coinbase transaction; for all others, we attributed each block to a single node.\footnote{The coinbase transaction is the first transaction in a block on a blockchain network such as Bitcoin, not to be confused with the eponymous cryptocurrency exchange Coinbase. This transaction is unique because it creates new coins and awards them to the miner who successfully added the block to the blockchain. It also typically includes the distribution of transaction fees collected from other transactions in the block, which are likewise awarded to the miner. In the context of mining pools, the coinbase transaction distributes these rewards proportionally among all pool participants, based on their contributed computational power.} For Ethereum, we correct the identification of validators for proposer-builder separation (PBS), as explained in Section~\ref{sec_appendix_mev}, thus fully accounting for the coincidence of the implementation of MEV-Boost on the same day as the Ethereum Merge. Further, we report the prevalence of MEV blocks and test for robustness in Section~\ref{sec_appendix_postpbs}. In PBS, the \textit{builder} of a block, who orders the transactions within the block, first receives the block reward and then transfers the block reward to the \textit{proposer} of the block, which is another name for the validator. Thus, one cannot simply use the block reward recipient data contained in the block and must find the proposer within the transactions of the block.


\begin{table}[ht]
\centering
\resizebox{\textwidth}{!}{
\begin{threeparttable}
\begin{tabular}{cccccc}
\hline \\[-1.8ex] 
\textbf{Blockchain}&  \textbf{Permissionless} & \textbf{Start date} & \textbf{Block time} & \textbf{Number of blocks} & \textbf{Number of nodes} \\
\hline \\[-1.8ex] 
Bitcoin & yes & Jan 3, 2009 & 10 min & 2,359,386 & 84,017 \\\\
Ethereum & yes & July 30, 2015 & 13 sec & 17,930,827 & 17,680 \\\\
Solana & yes & Mar 16, 2020 & 800 ms & 31,372,393 & 3,363 \\\\
BNB & no & Aug 29, 2020 & 5 sec & 31,372,303 & 60 \\\\
Gnosis & yes\textsuperscript{\textdagger} & Oct 8, 2018 & 3 sec & 30,315,947 & 1168 \\\\
Ronin & no\textsuperscript{\textdagger\textdagger} & Jan 25, 2021 & 3 sec & 22,973,856 & 39 \\\\
\hline 
\end{tabular}
\begin{tablenotes}
\small
\item \textsuperscript{\textdagger}Gnosis upgraded from Proof-of-Authority to Proof-of-Stake in December 2022.\\
\textsuperscript{\textdagger\textdagger}Ronin upgraded from a Proof-of-Authority to Delegated Proof-of-Stake in April 2023.
\end{tablenotes}
\end{threeparttable}
}
\vspace{1em}
\caption{Summary of blockchain data collected in the study} \label{tb_data}
\end{table}

\subsection{Variable Construction} \label{sec_variable}

\paragraph{Decentralization Variables:} We use multiple measures of decentralization for the robustness and interpretability of our estimated effects: Shannon entropy, the number of nodes, the Gini coefficient, the Nakamoto coefficient, and the Herfindahl–Hirschman index (HHI).

We define the measures as follows. The number of nodes, which we refer to as $\text{Nodes}_{it}$, is simply the number of nodes $N_{it}$ of blockchain $i$ that produced at least one block on day $t$. 

Shannon entropy, which we refer to as $\text{Entropy}_{it}$, is a measure of the unpredictability or randomness in the distribution of blocks among nodes. Formally, it is defined as
\begin{equation*}
\text{Entropy}_{it}=-\sum_{x \in X_{it}}p(x)\log p(x) \text{,}
\label{eq_entropy}
\end{equation*}
where $X_{it}$ is the set of the number of blocks produced by each node in blockchain $i$ on day $t$ \citep{lin2021measuring, jia2022measuring}. It has units of bits (as in \textit{gigabit}) and is bounded by $[0,\log_2(N)]$. A higher entropy means more decentralization. Importantly, Shannon entropy captures and can be decomposed into the number of nodes \textit{and} the distribution of block production over those nodes. The distribution of block production can subsequently be measured by the Gini coefficient, the Nakamoto coefficient, and HHI. Thus, Shannon entropy is a holistic measure of decentralization. An intuition for entropy is that any system with an entropy of $H$ (\textit{e.g.}, an entropy of 8) is equivalent in measure to a system with $2^H$ equally contributing entities (\textit{e.g.}, 256 nodes with the same number of blocks validated in a day). 

We additionally use the Gini coefficient ($\text{Gini}_{it}$), which measures inequality in block production across nodes (bounded by $[0,1]$; higher means less decentralization); the Nakamoto coefficient ($\text{Nakamoto}_{it}$), which measures the minimum number of nodes required to collectively control over 51\% of block production \citep{srinivasan_lee_2017} (bounded by $[1,N]$; higher means more decentralization);\footnote{The threshold to take control over the consensus layer varies by blockchain. For example, Bitcoin requires 51\% of the network, while Ethereum and Solana require 66.7\%. To simplify our comparative analysis, we used 51\% as the threshold for the Nakamoto coefficient.} and the Herfindahl--Hirschman Index ($\text{HHI}_{it}$), which measures block production concentration (bounded by $[1/N,1]$; higher means less decentralization). Formal definitions are provided in Section~\ref{sec_appendix_measures} of the Appendix.

\paragraph{Treated Blockchain Variable:} The treated blockchain variable $\text{Chain}_{i}$ is a binary indicator variable for blockchain $i$. $\text{Chain}_{i}$ equals one if the observation is from the blockchain that receives the treatment or zero if the observation is from the compared blockchain that does not. For example, in our analysis of China's mining ban, $\text{Chain}_\text{Bitcoin}=1$, and $\text{Chain}_\text{Ethereum}=0$. For simplicity, the tables display $\text{Chain}_\text{Bitcoin}$ as $\text{Bitcoin}$ for China's mining ban and $\text{Chain}_\text{Solana}$ as $\text{Solana}$ for Hetzner's shutdown.

\paragraph{Post-Treatment Variable:} The post-treatment variable $\text{After}_t$ is a binary indicator variable for date $t$. It is zero if the observation is before the treatment date and one if the observation is on or after the treatment date.

\paragraph{Treatment Variable:} The variable $\text{Treatment}_{it}$ is the combination of both the post-shock and blockchain variables. It is formulated as the cross-term $(\text{After}_t \times \text{Chain}_i)$ and is shown as $\text{Treatment}_{it}$ for convenience and clarity. This variable is used to estimate the average treatment effects on the treated (ATT) $\delta$. For example, in our analysis of China's mining ban, the coefficient $\delta$ of the treatment variable represents the difference in Bitcoin's and Ethereum's entropy after the ban.

\begin{figure}[!ht]
\centering
{\includegraphics[width=0.55\linewidth]{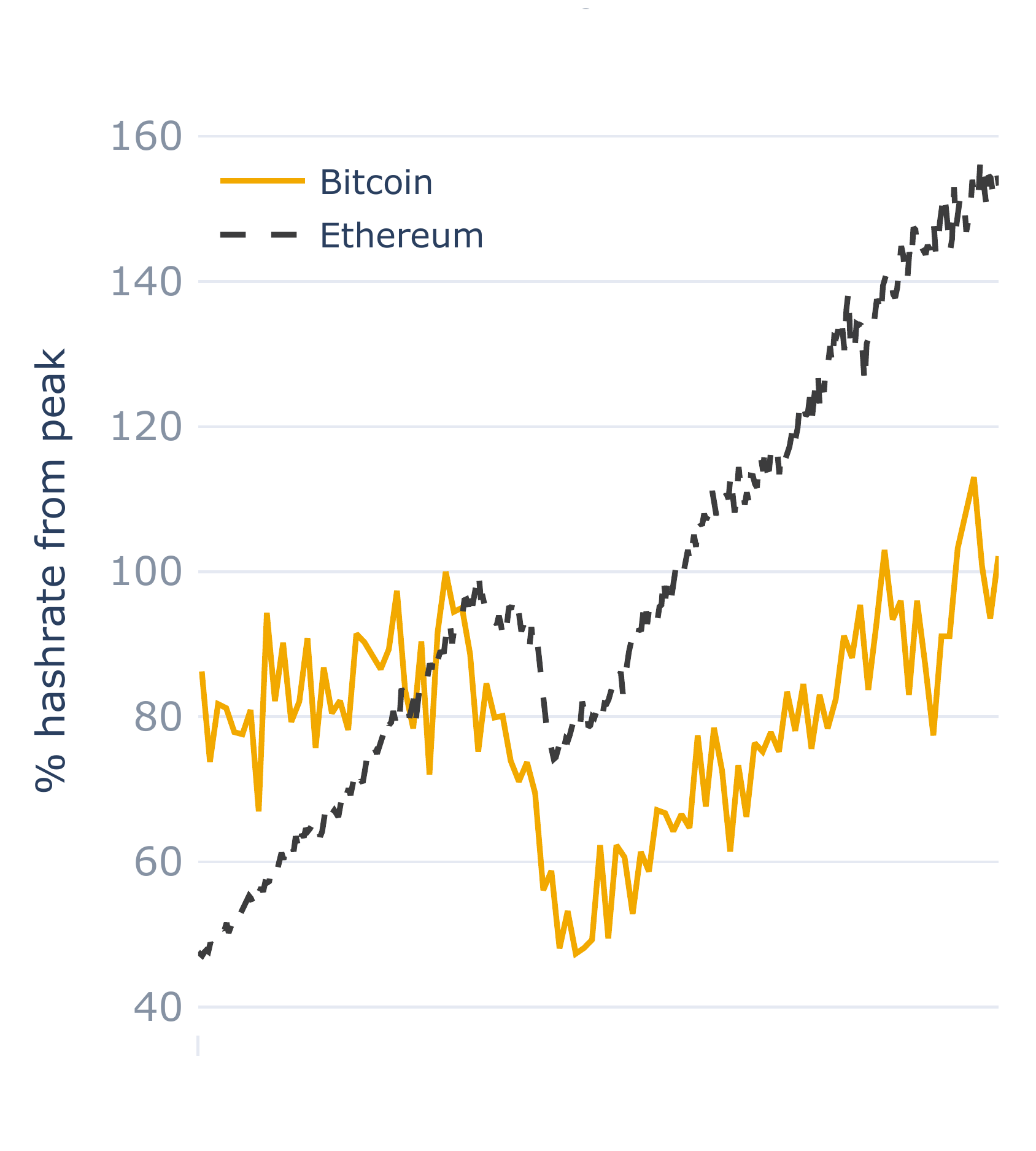}}
\caption{\small Percentage hashrates for Bitcoin and Ethereum from the local peak around China's ban on crypto mining.} 
\label{fig_hashrate}
\end{figure}

\paragraph{Exposure Variable:} China's mining ban disproportionately impacted Bitcoin's network compared to Ethereum's, as evidenced by the drop in hashrates starting at the time of the ban (Figure~\ref{fig_hashrate}). To control for the difference in exposure to the shock, we included the variable $\text{Exposure}_{it}$ as a covariate in our analysis of China's mining ban. It is formulated as the cross-term $(\text{After}_t \times \text{Exposure}_i)$, where $\text{Exposure}_i$ is the fractional decrease in the hashrate of blockchain $i$ from its peak hashrate at around the time of the ban to the minimum hashrate in the months following the ban.\footnote{Because the exposure lasts for months after the peak, we also perform multi-period difference-in-difference estimation in Section~\ref{sec_appendix_multiperiod} of the Appendix \citep{callaway2021diff}.} 

Non-random exposure to exogenous shocks can bias estimates \citep{borusyak2020nonrandom}. We directly measure exposure for both blockchains as the fractional drawdown of the total daily hashrate from its peak around May 15\textsuperscript{th} (Figure~\ref{fig_hashrate}), yielding $\text{Exposure}_\text{Bitcoin}=0.511$ and $\text{Exposure}_\text{Ethereum}=0.258$. These values reflect the differential concentration of mining in China before the ban: approximately 75\% of Bitcoin hashrate versus 25\% of Ethereum hashrate was estimated to be in China. By controlling for this differential exposure, our identification strategy isolates how each blockchain recovered from similar proportional shocks, which we attribute to differences in resource flexibility. We use the date of the peak hashrate for Bitcoin on May 15, 2021, as the date for the shock. As a robustness check, we validate our exposure measures using country-level geographic monthly hashrates for Bitcoin, serving as a proxy for both chains, in Section~\ref{sec_appendix_hashrate} of the Appendix.

\paragraph{Time Variable:} The time variable $\text{Day}_t$ is a running variable used in our event study. It is the number of days relative to the treatment date, taking negative values before treatment (\textit{e.g.}, $\text{Day}_t = -30$ for 30 days before), zero at the treatment date, and positive values after treatment.

\subsection{Model Specifications}

To estimate treatment effects in our analysis of the three shocks, we employ three empirical methods: difference-in-difference estimation (DiD), synthetic difference-in-difference (SDiD), and an event study.

\paragraph{Difference-in-Difference (DiD) Estimation} \label{sec_model_did}

We use DiD for the policy shock of China's mining ban and the infrastructure shock of Hetzner's shutdown of Solana nodes. The DiD model was specified as follows:
\begin{equation} \label{eq_did}
\begin{split}
\text{Y}_{it} &= \delta \text{Treatment}_{it} + \beta_1 \text{Chain}_{i} + \beta_2 \text{After}_t + \epsilon_{it}\text{,}
\end{split}
\end{equation}
where $\text{Y}_{it}$ is a measure of decentralization (\textit{i.e.}, $\text{Entropy}_{it}$, $\text{Nodes}_{it}$, $\text{Gini}_{it}$, $\text{Nakamoto}_{it}$, or $\text{HHI}_{it}$) of blockchain $i$ on day $t$ and $\epsilon_{it}$ is the error term. Including the exposure variable as a covariate yields:
\begin{equation} \label{eq_did_exp}
\begin{split}
\text{Y}_{it} &= \delta \text{Treatment}_{it} + \beta_1 \text{Chain}_{i} + \beta_2 \text{After}_t + \beta_3 \text{Exposure}_{it} + \epsilon_{it}\text{.}
\end{split}
\end{equation}
For China's ban, the variable $\text{Chain}_i$ is one for Bitcoin and zero for Ethereum. The variable $\text{Treatment}_{it}$ is the cross-term $(\text{Bitcoin} \times \text{After}_t)$ and indicates one for Bitcoin after the ban or zero for Bitcoin before the ban and for Ethereum before and after the ban. For Hetzner's shutdown of Solana nodes, the variable $\text{Chain}_i$ is one for Solana or zero for the synthetic control, which is explained below. The variable $\text{Treatment}_{it}$ indicates one for Solana after the shutdown or zero for Solana before the shutdown and for the synthetic control before and after the shutdown, or simply $(\text{Solana} \times \text{After}_t)$. The coefficient $\delta$ of the treatment variable $\text{Treatment}_{it}$ represents the difference in the treated chain and control chain following the shock.\footnote{For China's ban on mining, we perform multi-period difference-in-difference analysis in Section~\ref{sec_appendix_multiperiod} of the Appendix to account for the months-long exposure to the shock \citep{callaway2021diff}.}

In addition to estimating the treatment effects before and after the shocks, we aimed to estimate the dynamics of decentralization after the shocks by identifying the leading and lagging treatment effects. In our context, this analysis reveals if the blockchain recovered from the shock and, if so, how quickly. Thus, we modified Equation \ref{eq_did} by summing over the time-shifted treatment effects as specified by
\begin{equation} \label{eq_did_shift}
\begin{split}
\text{Y}_{it} &= \sum_{\lambda=-T}^T \delta_\lambda \text{Treatment}_{it\lambda} + \beta_1 \text{Chain}_{i} + \beta_2 \text{After}_t + \epsilon_{it}\text{,}
\end{split}
\end{equation}
where $\lambda$ is the lead and lag effects for $\lambda<0$ and $\lambda>0$, respectively and $T$ is the maximum lead and lag.

\paragraph{Synthetic Difference-in-Difference (SDiD) Estimation} \label{sec_model_sdid}

For the infrastructure shock of Hetzner's shutdown of Solana nodes, Solana exhibited no comparable control due to its unique entropy measures, which were significantly higher than those of other blockchains. Thus, we employ a synthetic difference-in-difference (SDiD) estimation approach using a synthetic control \citep{arkhangelsky2021synthetic}. This control is constructed as a weighted sum of the entropy values from four other blockchains: Ethereum, Gnosis, BNB, and Ronin. While these blockchains vary in their consensus mechanisms during the analysis period (Ethereum had recently transitioned to PoS, Gnosis used PoA before transitioning to PoS in December 2022, and BNB and Ronin used delegated or authority-based validation), all represent alternative blockchain networks that were \textit{not} affected by the Hetzner shutdown and thus serve as plausible counterfactuals for Solana's trajectory. The synthetic DiD method assigns weights to donor units based on pre-treatment fit \citep{arkhangelsky2021synthetic}, so chains that poorly match Solana's pre-treatment dynamics receive low weight, mitigating concerns about donor pool composition. The estimated weights are Gnosis (0.284), Ronin (0.282), BNB (0.280), and Ethereum (0.154), indicating that the four donor chains contribute roughly equally to the synthetic control.

The SDiD model is specified as follows:
\begin{equation} \label{eq_sdid}
\text{Y}_{it} = \tau_{it} \text{Treatment}_{it} + \beta_1 \text{Chain}_{i} + \epsilon_{it} \text{,}
\end{equation}
where $\text{Y}_{it}$ represents the decentralization of blockchain $i$ at time $t$ (see Section~\ref{sec_variable}), $\tau_{it}$ is the treatment effect of Hetzner's shutdown on the unit $i$ at time $t$, $\text{Treatment}_{it}$ is a binary indicator which equals 1 for Solana post-shutdown and 0 otherwise, $\text{Chain}_i$ is the blockchain fixed effects, and $\epsilon_{it}$ is the error term. To estimate the average treatment effect $\tau$ where the treatment was applied, we calculate $\tau$ as the average of $\tau_{it}$ over the observations with $\text{Treatment}_{it}=1$. All treated units begin treatment simultaneously. We computed the synthetic DiD estimators using the R package \verb|synthdid| \citep{arkhangelsky2021synthetic}.

\paragraph{Event Study}

We use an event study to estimate the effect of the software shock of the Ethereum Merge and to test the robustness of the difference-in-difference estimation for Hetzner's shutdown. For the Ethereum Merge, a DiD design with Bitcoin as control is problematic: Bitcoin did not undergo a comparable technical transition, and the two chains differ fundamentally in baseline validator structure and ecosystem composition, making a parallel trends assumption difficult to justify. More broadly, we could not find parallel trends between the decentralization of Ethereum and other blockchains around the event (as this was an Ethereum-only shock). The Event Study model was specified as follows:
\begin{equation} \label{eq_event}
\begin{split}
\text{Y}_{t} &= \delta \text{After}_t + \beta_1 \text{Day}_t + \beta_2 (\text{After}_t \times \text{Day}_t) + \epsilon_t\text{,}
\end{split}
\end{equation}
where $\text{Y}_t$ is a decentralization metric, $\text{After}_t$ is a binary indicator equal to one after the event, and $\text{Day}_t$ is a continuous running variable measuring the number of days relative to the event date (defined in Section~\ref{sec_variable}). We omit the subscript $i$ because Equation~\ref{eq_event} is estimated for a single blockchain at a time. This is a segmented regression specification used in interrupted time series analysis \citep{wagner2002segmented, bernal2017its}. The coefficient $\delta$ captures the immediate level shift in decentralization at the event date. The coefficient $\beta_1$ captures the pre-event daily trend: the slope of decentralization with respect to time before the event. The coefficient $\beta_2$ captures the \textit{change} in this slope at the event, so that the post-event daily trend is $\beta_1 + \beta_2$. A significant $\beta_2$ indicates that the shock altered not only the level but also the trajectory of decentralization over time.

We use an \textit{event study} and not \textit{regression discontinuity in time} to implement this estimation strategy because the effect was not localized around a specific time frame. In our case, the running variable, time, only moves forward and does not allow for a local comparison around the event to isolate a treatment effect. The effect of the Ethereum Merge was not localized at a specific time frame immediately around the event but was persistent after the event. This made an event study more suitable than regression discontinuity in time for our analysis, as we are capturing both the immediate impact of the Ethereum Merge and the longer-term trends associated with the Ethereum Merge \citep{hausman2017rdit}.

\section{Results}

In this section, we present the empirical results examining the role of resource flexibility, proxied by differences in blockchain design, on the decentralization of blockchain consensus in response to three shocks: policy, infrastructure, and technical. For each shock, we first present model-free evidence highlighting the pre- and post-shock values of decentralization. Then, we present DiD, SDiD, and event study estimates of the impact of the shock on decentralization. As discussed in Section~\ref{sec_theory}, the resource flexibility framework predicts that systems with greater resource flexibility should recover more effectively from disruptions. Taken together, the dynamics of the shocks and the comparisons between the three shocks provide strongly suggestive evidence consistent with this prediction: resource flexibility plays a role in influencing blockchain decentralization.

\subsection{China Bans Crypto Mining} \label{sec_results_china}

\begin{figure}[!ht]
\centering
{\includegraphics[width=0.55\linewidth]{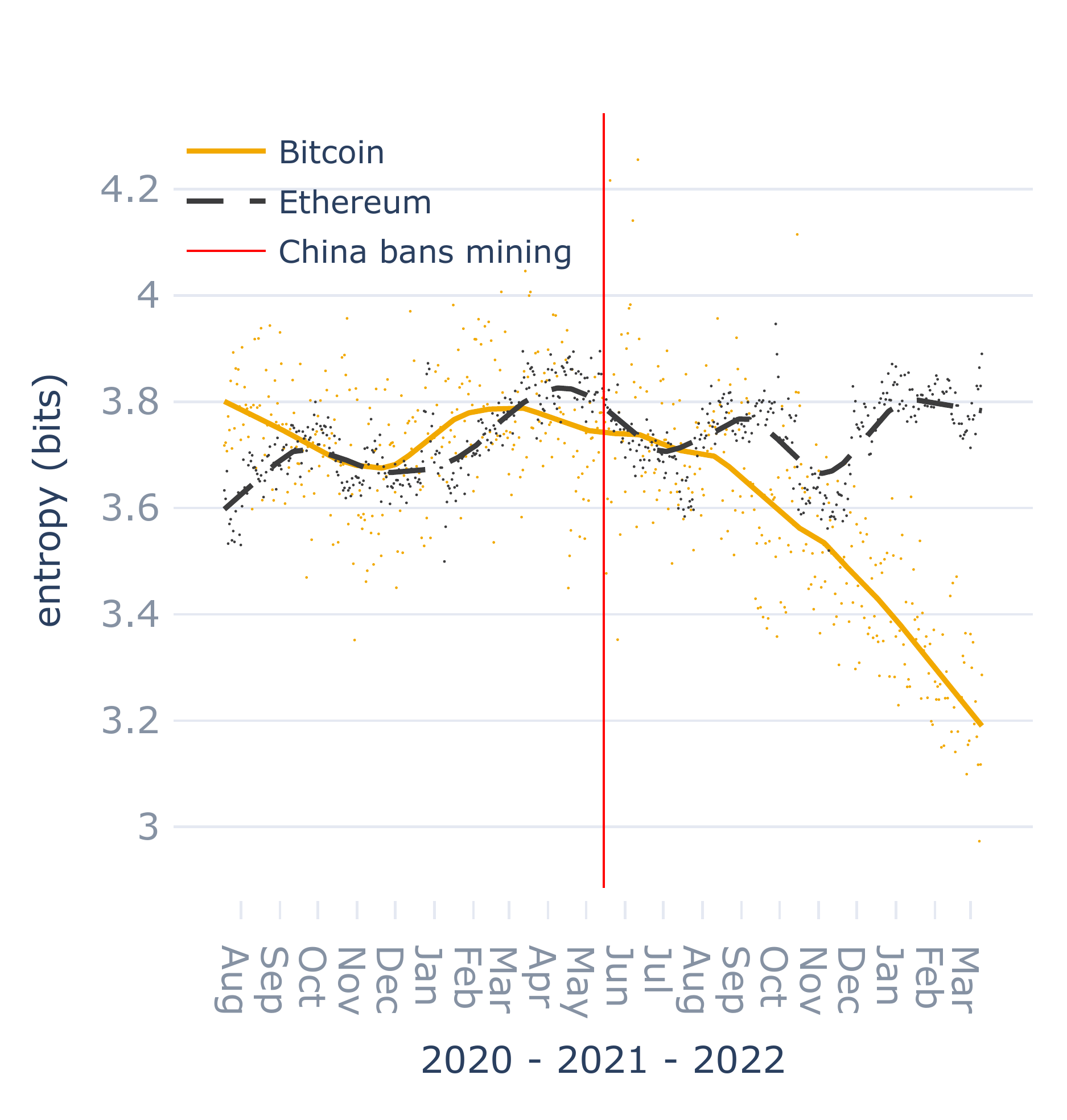}}
\caption{\small Decentralization (in bits) of Bitcoin and Ethereum around China's ban on crypto mining.} 
\label{fig_china}
\end{figure}

First, we study the effect of resource flexibility on the decentralization of the consensus of Bitcoin versus Ethereum in response to China's ban on crypto mining on May 15, 2021, as explained in Section~\ref{sec_experimental_setting}. Figure~\ref{fig_china} shows the decentralization of Bitcoin and Ethereum as measured by entropy in bits (as used to measure the quantity of digital information). Since our empirical analyses are based on DiD estimates, they rely on the parallel trends assumption. First, we show visual proof of the parallel trends assumption before the shock in Figure~\ref{fig_china}. While the entropy measures seem to display some seasonality (perhaps due to varying electricity costs for nodes across geographies), the decentralization metrics are similar in absolute terms as well as parallel in trends.

After China's ban on crypto mining, we observe a significant decline in entropy for Bitcoin over many months (Figure~\ref{fig_china}). In contrast, we see little to no decline in entropy for Ethereum. To estimate the effect, we analyzed the results of the DiD specifications in Equation~\ref{eq_did} to estimate the impact of resource flexibility on decentralization in response to the shock, by using the variation in hardware requirements as a proxy for resource flexibility, as explained in Section~\ref{sec_experimental_setting}. We denote $\text{Bitcoin} = \text{Chain}_{\text{Bitcoin}}$ for simplicity. In our analysis, we find that Bitcoin's entropy was reduced by 0.209 bits from 3.688 bits (by around 5.7\%) before the ban in comparison with Ethereum's entropy (Table~\ref{tb_china}). To further explore this effect, we present case studies in Section~\ref{sec_appendix_decompose} of the Appendix where we can observe individual nodes and mining pools known to be based in China that experience declines in block production.

\begin{table}[ht]
\centering
\resizebox{\textwidth}{!}{
\begin{threeparttable}
\begin{tabular}{lcccccccccc}
\hline \\[-1.8ex]
\textit{Dependent} & \multicolumn{2}{c}{\textit{Entropy}} & \multicolumn{2}{c}{\textit{Nodes}} & \multicolumn{2}{c}{\textit{Gini}} & \multicolumn{2}{c}{\textit{Nakamoto}} & \multicolumn{2}{c}{\textit{HHI}}\\
\textit{variable} & (1) & (2) & (3) & (4) & (5) & (6) & (7) & (8) & (9) & (10) \\
\hline \\[-1.8ex]
\textbf{Treatment}  & -0.209$^{***}$ & -0.201$^{***}$ & 5.050$^{***}$ & 4.943$^{***}$ & 0.047$^{***}$ & 0.047$^{***}$ & -0.414$^{***}$ & -0.404$^{***}$ & 0.012$^{***}$ & 0.012$^{***}$\\ 
 & (0.048) & (0.047) & (1.206) & (1.181) & (0.009) & (0.009) & (0.085) & (0.084) & (0.003) & (0.003)\\
\textbf{Exposure}  &  & -0.032 &  & 0.424 &  & 0.003 &  & -0.038 &  & 0.002\\ 
 &  & (0.021) &  & (0.576) &  & (0.007) &  & (0.041) &  & (0.001)\\
\textbf{Bitcoin}  & 0.020 & 0.020 & -33.980$^{***}$ & -33.980$^{***}$ & -0.327$^{***}$ & -0.327$^{***}$ & 1.577$^{***}$ & 1.577$^{***}$ & -0.037$^{***}$ & -0.037$^{***}$\\ 
 & (0.016) & (0.016) & (1.075) & (1.076) & (0.008) & (0.008) & (0.046) & (0.046) & (0.002) & (0.002)\\
\textbf{After}  & 0.064 & 0.072 & -3.093 & -3.203 & -0.033 & -0.034 & 0.241 & 0.251 & -0.004 & -0.004\\ 
 & (0.093) & (0.089) & (2.528) & (2.397) & (0.032) & (0.031) & (0.179) & (0.170) & (0.006) & (0.005)\\
\textbf{Intercept}  & 3.672$^{***}$ & 3.672$^{***}$ & 52.990$^{***}$ & 52.990$^{***}$ & 0.791$^{***}$ & 0.791$^{***}$ & 3.069$^{***}$ & 3.069$^{***}$ & 0.138$^{***}$ & 0.138$^{***}$\\ 
 & (0.065) & (0.065) & (0.593) & (0.594) & (0.014) & (0.014) & (0.206) & (0.206) & (0.009) & (0.009)\\
\hline \\[-1.8ex]
Monthly FE & Yes & Yes & Yes & Yes & Yes & Yes & Yes & Yes & Yes & Yes \\
Observations & 1202 & 1202 & 1202 & 1202 & 1202 & 1202 & 1202 & 1202 & 1202 & 1202 \\
\hline
\hline
\end{tabular}
\begin{tablenotes}
\small
\item Notes: $^{*}$p$<$0.05; $^{**}$p$<$0.01; $^{***}$p$<$0.001. Standard errors are clustered by blockchain and month, as clustering at the blockchain level alone yields only 2 clusters, which is insufficient for reliable asymptotic inference \citep{cameron2008bootstrap, mackinnon2023cluster}. See Section~\ref{sec_appendix_clustering} for a robustness check with blockchain-level clustering. Monthly fixed effects (FE) are included.
\end{tablenotes}
\end{threeparttable}
}
\vspace{1em}
\caption{China bans crypto mining} \label{tb_china}
\end{table}

It is important to note that Bitcoin and Ethereum experienced different exposures to the mining ban, which could have led to omitted variable bias \citep{borusyak2020nonrandom}. We therefore controlled for exposure by including the covariate $\text{Exposure}_{it}$, described by Equation~\ref{eq_did_exp}. Then, we observed that the magnitude of the effect size for $\text{Treatment}$ decreased slightly to $-0.201$ while the $\text{Exposure}$ covariate had a small, non-significant effect of $-0.032$ (Table~\ref{tb_china}, Column 2). The slight decrease in the effect size after controlling for exposure underscores the robustness of our findings, suggesting that the observed decline in Bitcoin's decentralization is not solely attributable to differences in exposure to the mining ban but also likely influenced by the underlying resource flexibility of the blockchain's consensus mechanism.

We also estimated the effects for the other metrics (Table~\ref{tb_china}). Surprisingly, we observed that the number of nodes increased for Bitcoin after the shock by around 5 nodes.\footnote{This positive effect on the number of nodes is possibly due to the dissolution of mining pools in China due to the ban. See Section~\ref{sec_appendix_decompose} for case studies of China's mining pools.} Despite the increase in the number of nodes, all other metrics, which measure the \textit{distribution} of block production among nodes, indicate that Bitcoin became more centralized after the shock. The Gini coefficient increased by 0.047, the Nakamoto coefficient decreased by around 0.414, and HHI increased by 0.012, all of which indicate the post-shock centralization of Bitcoin.\footnote{Note that $\text{Gini}$ and $\text{HHI}$ are bounded between 0 and 1. The units are in bits for $\text{Entropy}$ and nodes for $\text{Nodes}$ and $\text{Nakamoto}$.} Taken together, the results suggest that the addition of around 5 nodes were nodes with low block production and overall show a significant negative impact of the mining ban on the decentralization of Bitcoin.

\subsection{Hetzner Shuts Down Solana Validators} \label{sec_results_hetzner}

\begin{figure}[!ht]
\centering
{\includegraphics[width=0.55\linewidth]{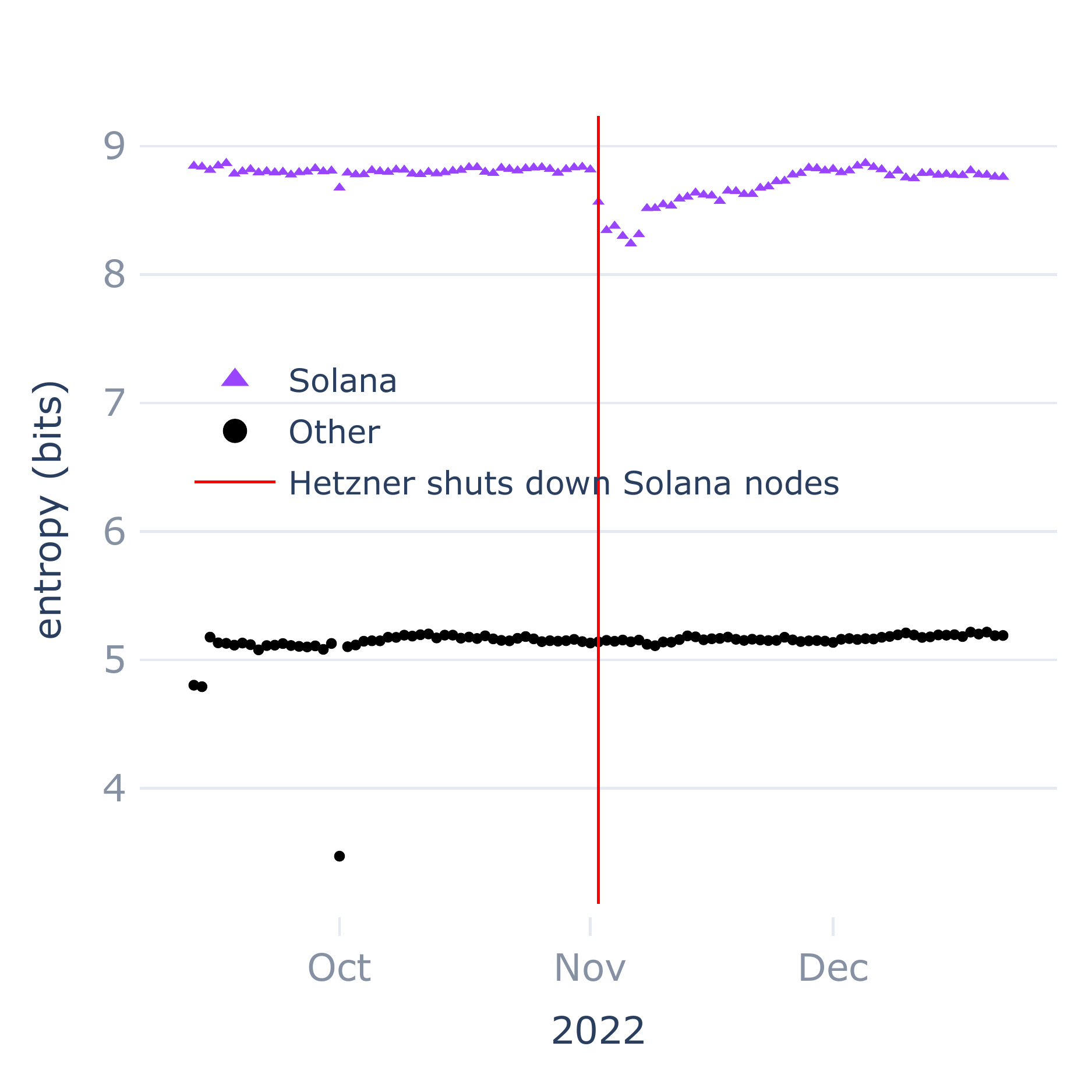}}
\caption{\small Decentralization (in bits) of Solana and a synthetic control around Hetzner's shutdown of Solana validators.}
\label{fig_hetzner}
\end{figure}

Second, we study the effect of Hetzner shutting down Solana validators on November 2, 2022, on the decentralization of Solana's consensus mechanism, as explained in Section~\ref{sec_experimental_setting}. We compare Solana to a synthetic control constructed from a weighted average of the other blockchains which can all experience infrastructure shocks. While this analysis examines a single blockchain rather than comparing across resource flexibility levels, its value lies in the comparison with the China ban: both represent ``forced relocation'' shocks where consensus participants were suddenly unable to operate from their existing infrastructure. Comparing recovery dynamics across these shocks (hardware-based Bitcoin versus token-based Solana) reveals how resource flexibility affects resilience. Figure~\ref{fig_hetzner} shows the entropy values of Solana and the synthetic control. Before estimating DiD effects, we visually demonstrate the parallel trends assumption between Solana and the synthetic control by showing the daily entropy of the blockchains over the period before the shock. 

On the day of and following Hetzner's shutdown of Solana validators, we observe an immediate, visible drop in the entropy of Solana. By comparison, we saw no shock to the entropy of the synthetic control. To estimate this effect, we estimate the synthetic difference-in-difference model in Equation~\ref{eq_sdid} with Solana and a synthetic control of four other blockchains (Ethereum, Gnosis, BNB, and Ronin). We denote $\text{Chain}_{\text{Solana}}$ as $\text{Solana}$ for clarity. We find that the shutdown reduced 0.378 bits of entropy (around 4.30\% from a pre-shock average of 8.787) of Solana ten days before and after the shock, in comparison with the synthetic control (Table~\ref{tb_hetzner}, Column 1). Then having decomposed the entropy measure into the nodes and measures of the distribution of block production across nodes, we find that around 326 nodes went offline in this period  (Table~\ref{tb_hetzner}, Column 2). This decline was accompanied by an increase in the Gini coefficient of 0.041 and a decrease in the Nakamoto coefficient of -6.202 nodes but no significant change in HHI  (Table~\ref{tb_hetzner}, Columns 3, 4, and 5). These findings indicate a substantial negative effect of the infrastructure disruption on the decentralization of Solana.

\begin{table}[ht]
\centering
\begin{threeparttable}
\begin{tabular}{lccccc} 
\hline \hline \\[-1.8ex] 
\textit{Dependent} & \textit{Entropy} & \textit{Nodes} & \textit{Gini} & \textit{Nakamoto} & \textit{HHI}\\
\textit{variable} & (1) & (2) & (3) & (4) & (5) \\
\hline
\textbf{Treatment}  & -0.378$^{***}$ & -326.120$^{***}$ & 0.041$^{***}$ & -6.202$^{***}$ & 0.001$^{***}$\\
 & (0.001)& (3.683)& (0.006)& (0.078)& (0.000)\\

\hline \\[-1.8ex]
Observations        & 42        & 42        & 42       & 42       & 42 \\ 
\hline
\hline 
\end{tabular}
\begin{tablenotes}
\small
\item Notes: $^{*}$p$<$0.05; $^{**}$p$<$0.01; $^{***}$p$<$0.001. Standard errors are derived from placebo tests.
\end{tablenotes}
\end{threeparttable}
\vspace{1em}
\caption{Hetzner shuts down Solana nodes.} \label{tb_hetzner}
\end{table}

\begin{figure}[!ht]
\centering
{\includegraphics[width=\linewidth]{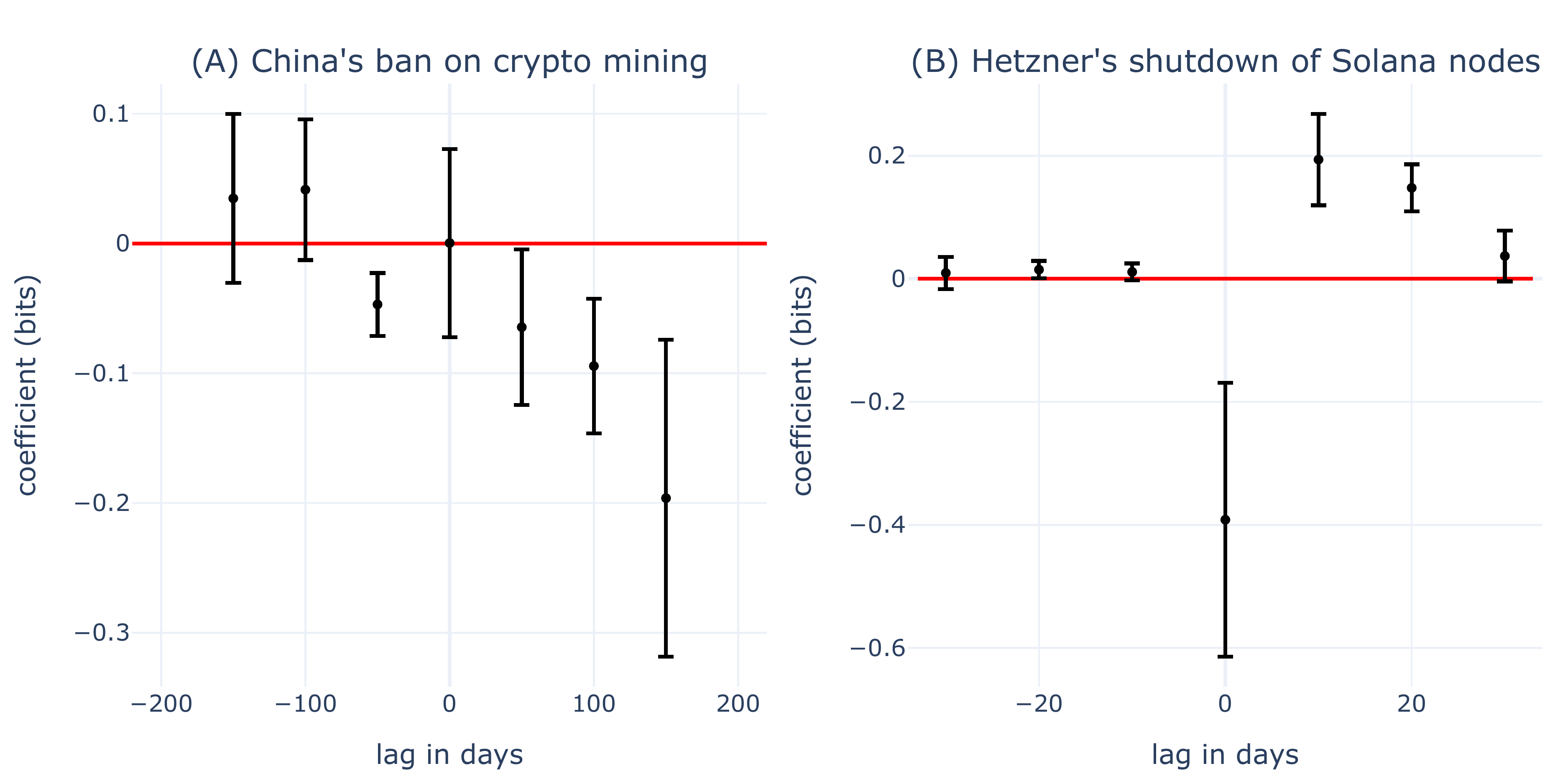}}
\caption{\small Coefficients for the lead and lag effects $\text{Treatment}_\lambda$ on the entropy of (A) China's mining ban and (B) Hetzner's shutdown of Solana validators for $\lambda$ lags in days. Error bars indicate 95\% confidence intervals.} 
\label{fig_lag_comparison}
\end{figure}

To compare Hetzner's shutdown to China's mining ban, we estimate lagged effects using Equation~\ref{eq_did_shift}.\footnote{Note that we used blockchain fixed effects instead of a synthetic control in SDiD since we wanted to estimate multiple treatment effects, unlike in Table~\ref{tb_hetzner}.} In Figure~\ref{fig_lag_comparison}, the Hetzner shutdown reduced Solana's entropy by 0.392 bits immediately after the shock, but entropy increased in the 40 days following the shock.\footnote{Note that the post-shock positive lagged effects in Figure~\ref{fig_lag_comparison}B are measuring the increase in entropy following the initial large dip in response to the shock; Table~\ref{tb_hetzner} shows the average post-shock effect.} In contrast, the treatment lag for the China ban lasts for over 180 days after the shock. The coefficients for all metrics are displayed in Sections~\ref{sec_appendix_lagged_china} and \ref{sec_appendix_lagged_hetzner} of the Appendix. Moreover, for the Hetzner shutdown, we estimate the DiD specification at $\pm$10 to $\pm$50 days in Section~\ref{sec_appendix_hetzner_bandwidth} of the Appendix; the effect at $\pm$50 days is not statistically significant and only 63\% of the effect at $\pm$10 days.

To further test robustness, we estimate Equation~\ref{eq_event} of Solana's entropy as an event study which revealed results consistent with our difference-in-difference estimation. Our results, shown in Table~\ref{tb_hetzner_event}, indicate that the shutdown had a negative effect of 0.291 bits on entropy at the time of the shock. All other metrics also show centralizing effects after the shutdown, with the number of nodes decreasing by around 337, the Gini coefficient increasing by 0.016, the Nakamoto coefficient decreasing by 6.637, and HHI increasing by 0.001. Then, the change in slope ($\hat{\beta}_2 = 0.006^{***}$) indicates that entropy recovered at 0.006 bits per day beyond the negligible pre-shock trend ($\hat{\beta}_1 \approx 0$). Dividing the initial level shift ($\hat{\delta} = 0.291$ bits) by this recovery rate yields an estimated recovery time of approximately 49 days. These results are again consistent with our DiD estimation.

\begin{table}[htbp]
\centering
\begin{threeparttable}
\begin{tabular}{lccccc} 
\hline \hline \\[-1.8ex] 
\textit{Dependent variable} & \textit{Entropy} & \textit{Nodes} & \textit{Gini} & \textit{Nakamoto} & \textit{HHI} \\
\hline \\[-1.8ex] 
\textbf{After}  & -0.291$^{***}$ & -336.966$^{***}$ & 0.016$^{**}$ & -6.637$^{***}$ & 0.001$^{***}$\\ 
 & (0.033) & (27.514) & (0.006) & (0.574) & (0.000)\\
\textbf{Days}  & -0.000$^{*}$ & 3.505$^{***}$ & 0.000$^{***}$ & -0.016$^{*}$ & 0.000$^{***}$\\ 
 & (0.000) & (0.070) & (0.000) & (0.007) & (0.000)\\
\textbf{Days $\times$ After}  & 0.006$^{***}$ & -1.656$^{*}$ & -0.001$^{***}$ & 0.217$^{***}$ & -0.000$^{***}$\\ 
 & (0.001) & (0.678) & (0.000) & (0.015) & (0.000)\\
\textbf{Intercept}  & 8.802$^{***}$ & 2063.742$^{***}$ & 0.723$^{***}$ & 55.467$^{***}$ & 0.006$^{***}$\\ 
 & (0.006) & (2.566) & (0.001) & (0.236) & (0.000)\\
\hline \\[-1.8ex] 
Observations                    & 121 & 121 & 121 & 121 & 121 \\
\hline
\hline 
\end{tabular}
\begin{tablenotes}
\small
\item Notes: $^{*}$p$<$0.05; $^{**}$p$<$0.01; $^{***}$p$<$0.001. Heteroskedasticity-robust standard errors.
\end{tablenotes}
\end{threeparttable}
\vspace{1em}
\caption{Event study on Hetzner shutting down Solana nodes} \label{tb_hetzner_event}
\end{table}

The difference in the dynamics of China's ban and Hetzner's shutdown can be attributed to two key factors: the first in the nature of the shock and the second in the design of the blockchains. First, China's ban was a policy ban while Hetzner's shutdown was an infrastructure shutdown. Policy changes take time to comply with and enforce, especially for distributed networks such as Bitcoin.\footnote{We show more evidence of this rolling ban of Bitcoin miners in Sections~\ref{sec_appendix_multiperiod}, \ref{sec_appendix_hashrate}, and \ref{sec_appendix_decompose} of the Appendix.} In contrast, Hetzner themselves locked the servers that were validating for Solana.\footnote{\label{fn_hetzner_tweets}See tweets of Hetzner's notice by \href{https://twitter.com/foldfinance/status/1587934819245207553}{@foldfinance} and \href{https://twitter.com/solblaze_org/status/1587818136694591491}{@solblaze\_org}. Accessed August 23, 2023.} So the impact of an infrastructure shock may be more immediate, as we have seen in the data. The two shocks also differed in magnitude: Hetzner's shutdown affected approximately 20\% of Solana's staked validators, while China's ban displaced roughly 75\% of Bitcoin's hashrate. While this severity difference contributes to the difference in recovery timelines, it cannot account for the within-shock comparison: under the same China ban, Ethereum (with more flexible GPUs) experienced minimal decline while Bitcoin (with inflexible ASICs) suffered lasting decentralization loss.

Second, Bitcoin and Solana differ in their consensus mechanisms. Bitcoin uses specialized hardware (i.e., application-specific integrated circuits \textit{ASICs}) and energy-intensive computation to achieve PoW consensus. ASICs are, by definition, specific to Bitcoin mining and are not readily available in other cloud facilities. By comparison, Solana uses staked native tokens, SOL, to achieve PoS consensus. In Solana, one can relatively easily create new validators by transferring their tokens to another cloud server or even a home server.\footnoteref{fn_hetzner_tweets} Our results thus show that the flexibility of resources used in the consensus mechanism (i.e., tokens versus hardware) appears to have reduced frictions in recovering from shocks, suggesting a role in influencing sustained decentralization.

\subsection{The Ethereum Merge} \label{sec_results_merge}

\begin{figure}[!ht]
\centering
{\includegraphics[width=0.55\linewidth]{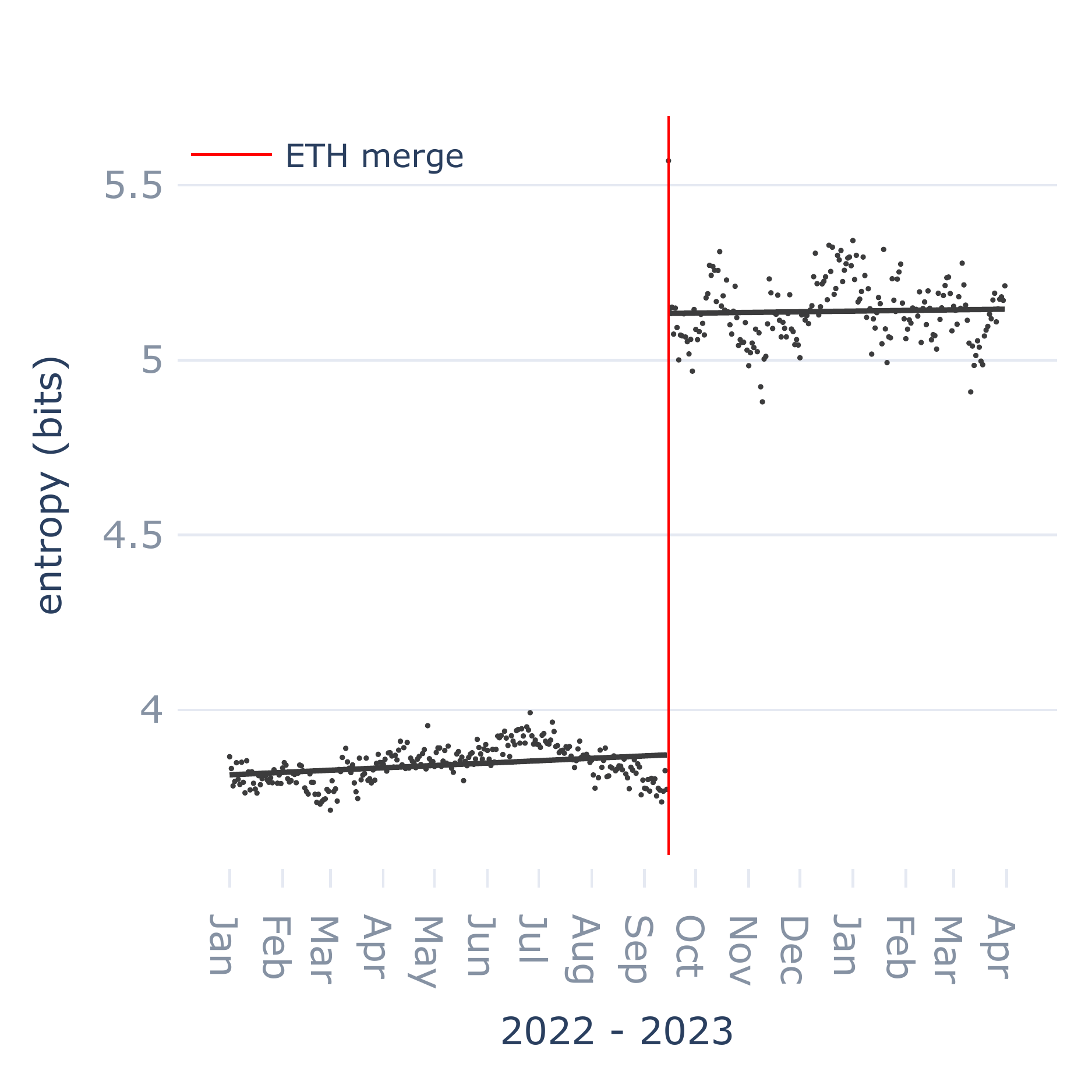}}
\caption{\small Daily decentralization (in bits) of Ethereum's consensus around the Ethereum Merge.} \label{fig_merge}
\end{figure}

In the two previous natural experiments, we examine the effects of resource flexibility on decentralization in response to shocks across blockchains that differed in their level of resource flexibility. In this section, we investigate the effect of a technical upgrade, the Ethereum Merge on September 15, on the decentralization of a single blockchain over time, as explained in Section~\ref{sec_experimental_setting}. The Merge changed Ethereum's consensus from PoW to PoS, removing the need for physical hardware and allowing anyone with 32 ETH and consumer-grade computers to participate. By replacing hardware requirements (expensive GPUs, cheap electricity, technical expertise) with token requirements, the Merge directly lowered barriers to entry, the second mechanism through which the resource flexibility framework predicts increased decentralization (Section~\ref{sec_theory}). Importantly, while the Merge also improved speed, security, and energy efficiency, these improvements do not mechanically predict increased decentralization, as efficiency gains often favor large operators through economies of scale. The resource flexibility mechanism is distinct: the specific pattern across our five metrics (increased entropy and nodes, increased Gini, unchanged Nakamoto and HHI) is uniquely consistent with a barrier-reduction mechanism rather than generic performance improvements. Figure~\ref{fig_merge} shows the entropy of Ethereum's consensus before and after the Merge, where we can visually observe the stepwise increase in both the means and the variance of decentralization.

To estimate this impact on Ethereum's decentralization, we estimate the specification for the Event Study in Equation~\ref{eq_event}. We find that the Merge increased Ethereum's decentralization by 1.27 bits of entropy from 3.872 bits (Table~\ref{tb_merge}). This increase in entropy can be attributed to the addition of around 680 new nodes, direct evidence of the barrier-reduction mechanism predicted by the resource flexibility framework (Section~\ref{sec_theory}). The accompanying increase in the Gini coefficient of 0.124 indicates that these new entrants are predominantly smaller validators, which increases measured inequality while improving overall decentralization through broader participation. Interestingly, the analyses show no significant change in the Nakamoto coefficient or the HHI. This implies that the number of nodes that control 51\% of the network remained similar despite the increase in the broader base of the network, consisting mainly of smaller new entrants. This pattern indicates that resource flexibility primarily operates at the extensive margin by broadening the base of participants rather than redistributing power among the largest validators. Affecting concentration at the intensive margin likely requires complementary mechanisms beyond resource flexibility alone.

\begin{table}[ht]
\centering
\begin{threeparttable}
\begin{tabular}{lccccc} 
\hline \hline \\[-1.8ex] 
\textit{Dependent variable}     & \textit{Entropy} & \textit{Nodes} & \textit{Gini} & \textit{Nakamoto} & \textit{HHI} \\
\hline \\[-1.8ex] 
\textbf{After}  & 1.262$^{***}$ & 693.574$^{***}$ & 0.124$^{***}$ & 1.312$^{***}$ & -0.035$^{***}$\\ 
 & (0.037) & (13.927) & (0.006) & (0.179) & (0.004)\\
\textbf{Days}  & 0.000 & 0.023$^{***}$ & 0.000$^{***}$ & 0.000 & -0.000$^{**}$\\ 
 & (0.000) & (0.004) & (0.000) & (0.000) & (0.000)\\
\textbf{Days $\times$ After}  & -0.000 & 0.295 & -0.000 & 0.003$^{*}$ & -0.000\\ 
 & (0.000) & (0.155) & (0.000) & (0.001) & (0.000)\\
\textbf{Intercept}  & 3.872$^{***}$ & 52.104$^{***}$ & 0.774$^{***}$ & 3.028$^{***}$ & 0.122$^{***}$\\ 
 & (0.030) & (0.617) & (0.002) & (0.026) & (0.003)\\
\hline \\[-1.8ex] 
Observations                    & 455 & 455 & 455 & 455 & 455 \\
\hline
\hline 
\end{tabular}
\begin{tablenotes}
\small
\item Notes: $^{*}$p$<$0.05; $^{**}$p$<$0.01; $^{***}$p$<$0.001. Standard errors are clustered by month to account for within-month serial correlation in daily observations \citep{cameron2008bootstrap}. With approximately 15 months of data, there are sufficient clusters for reliable inference. The shorter Hetzner event study window (Table~\ref{tb_hetzner_event}) uses heteroskedasticity-robust standard errors instead, as 4 months would yield too few clusters.
\end{tablenotes}
\end{threeparttable}
\vspace{1em}
\caption{The Ethereum Merge} \label{tb_merge}
\end{table}

\section{Discussion and Conclusions} 

We use data from major blockchain networks to investigate the impact of significant shocks to blockchain consensus decentralization. Specifically, we analyze three policy-related, infrastructure-related, and technical shocks in the blockchain industry, and our findings strongly suggest that resource flexibility plays a significant role in the decentralization dynamics of blockchains and enables sustained blockchain consensus decentralization across the blockchains we study. These findings are consistent with the resource flexibility framework (Section~\ref{sec_theory}): systems whose consensus resources have lower asset specificity can reallocate more flexibly and recover faster from disruptions. As Figure~\ref{fig_lag_comparison} illustrates, Bitcoin's reliance on specialized hardware for PoW consensus made it more vulnerable to policy shocks, whereas Solana's PoS mechanism allowed for a more agile response to infrastructure disruptions.

\subsection{Implications for Blockchains}

Our study provides empirical insights into the factors affecting blockchain decentralization, which informs various stakeholders in the blockchain ecosystem. First, the data suggests that design elements emphasizing resource flexibility are associated with a blockchain's ability to maintain decentralization in the face of external shocks. However, this flexibility may come with trade-offs. More flexible resources may be easier for attackers to acquire, potentially enabling faster concentration of power for attacks, though our evidence shows flexibility also enables faster defensive recovery. Higher decentralization typically involves more nodes, which can reduce throughput and increase latency. And more decentralized networks face greater coordination costs for upgrades and governance decisions, as illustrated by the years of coordination required for the Ethereum Merge itself. Moreover, the optimal level of resource flexibility may depend on the threat environment: as \cite{garratt2023fixed} show, inflexibility can be protective when shocks target mining \textit{rewards}, whereas our results suggest that flexibility is protective when shocks target the \textit{resources} themselves. This \textit{specificity-flexibility tradeoff} reflects a general economic principle: asset specificity protects against demand-side shocks (sunk costs create exit barriers that maintain participation), while asset flexibility protects against supply-side shocks (portable resources enable relocation and redeployment) \citep{williamson1985}. Second, operational decisions appear to have a measurable impact on a blockchain's decentralization. These decisions could either enhance or compromise a blockchain's structural integrity, contingent on its specific design. Lastly, policy decisions targeting blockchains have the potential for broader, systemic effects. Therefore, if the goal is to maintain a balance between innovation and safety, policymakers should consider the unintended consequences of their regulations on the decentralization and overall stability of multiple blockchains.

\paragraph{Blockchain Design}

Our findings suggest that design choices emphasizing resource flexibility can improve resilience to external shocks. Alternative design approaches like stateless nodes and blobs could further increase participation by lowering hardware requirements \citep{rustgi2023future}.\footnote{See \href{https://ethereum.org/en/roadmap/statelessness/}{ethereum.org}'s explanation on statelessness and \href{https://eips.ethereum.org/EIPS/eip-4844}{EIP-4844} for details on blobs. Accessed April 15, 2024.} However, our analysis of the Merge shows that despite increases in entropy and participation, concentration among the largest validators remained unchanged, highlighting the ongoing challenges in achieving further decentralization beyond the extensive margin. Across all three shocks, we observe changes in the breadth of participation (entropy, node counts) more consistently than in the concentration of power among top validators (Nakamoto, HHI), suggesting that resource flexibility is a necessary but not sufficient condition for full decentralization.

\paragraph{Blockchain Operations}

Our findings indicate that reliance on a limited number of infrastructure providers could expose blockchains to vulnerabilities, as evidenced by the prolonged impact on Bitcoin's decentralization following China's ban \citep{dunbar2023evaluating}. This suggests that if resilience against regulatory changes is a priority, operators might consider diversifying their infrastructure across multiple jurisdictions. The recovery dynamics observed across shocks further point to the role of the consensus mechanism in determining resilience: operators of PoW blockchains must consider hardware supply and geographic concentration, while those managing PoS blockchains could focus on infrastructure provider diversification.\footnote{See the Solana Foundation's \href{https://solana.com/news/validator-health-report-march-2023}{March 2023 Report} and \href{https://nakaflow.io/}{nakaflow.io}. Accessed September 6, 2023.}

\paragraph{Blockchain Policy}

Our findings indicate that policy changes can significantly influence blockchain decentralization, as evidenced by the centralization of Bitcoin in response to China's ban (Figure~\ref{fig_china}, Table~\ref{tb_china}). Policymakers should be cognizant of the potential ripple effects on the blockchain ecosystem, particularly as an increasing number of financial assets, including securities and government bonds, are being recorded on blockchains \citep{eib2021digitalbond}. Understanding the implications of policy decisions on both the integrity and the decentralization of these platforms is increasingly important.

\subsection{Limitations and Future Research} \label{sec_limitations}

While our study offers insights into the dynamics of blockchain design, operations, and policy, there are some limitations to consider. Our biggest limitation is that our analyses are of natural experiments in which we cannot vary just the resource flexibility of different blockchains. To better understand this limitation, we consider alternative hypotheses to ours. First, the type of event varied across shocks, which include policy-related, infrastructure-related, and technical shocks. For example, the lasting effect of China's mining ban compared to the short-lived effect of Hetzner's shutdown could be because China's mining ban was a policy shock and Hetzner's shutdown was an infrastructure shock. Intuitively, policy changes take days to weeks to implement and months to enforce, whereas infrastructure shocks can occur within days. However, we observe that Ethereum experienced the same shock as Bitcoin, but that Ethereum, whose consensus mechanism was inherently more flexible, saw little to no negative effect on its decentralization. Moreover, we perform multi-period difference-in-difference estimation and node-level decomposition in Sections~\ref{sec_appendix_multiperiod} and \ref{sec_appendix_decompose} of the Appendix, respectively, to test for robustness of the results at finer resolutions of both time and units, further providing strongly suggestive evidence for our resource flexibility hypothesis.

Second, the reverse causality concern merits explicit discussion: could the level of decentralization determine resource flexibility, rather than the reverse? We address this on both theoretical and empirical grounds. Theoretically, resource flexibility in our setting is a design-time property of the consensus protocol, not an endogenous outcome. Bitcoin requires ASICs because of its SHA-256 hash function, chosen at launch in 2009; Ethereum required GPUs because of its Ethash algorithm, designed to resist ASIC optimization; and Solana requires staked tokens because of its Proof-of-Stake architecture. These resource requirements are specified in the protocol and cannot be altered by the current level of decentralization. Empirically, our identification strategy further mitigates this concern. The three shocks are exogenous to the blockchains' decentralization levels: China's ban targeted energy consumption, Hetzner enforced existing terms of service, and the Merge followed a multi-year technical roadmap. The Ethereum Merge provides a particularly clean test, as the same blockchain transitions from less flexible (PoW) to more flexible (PoS) resources through a planned upgrade determined years in advance, ruling out contemporaneous decentralization levels as a driver of the change in resource flexibility. While we cannot claim to have fully resolved all endogeneity concerns, the design-time nature of resource flexibility and the exogeneity of our shocks substantially limit the scope for reverse causality.

Third, positive market sentiment could facilitate quicker recovery by encouraging reinvestment and participation. To test this, we estimate Equation~\ref{eq_event} on token prices around each shock and find that all three had significant negative price effects: BTC dropped \$14,880 ($p\ll0.001$), SOL dropped \$7.25 ($p\ll0.001$), and ETH dropped \$351.51 ($p\ll0.001$), ruling out positive sentiment as a driver of the differential recovery dynamics (Section~\ref{sec_appendix_price}).

Fourth, Sybil attacks are possible on blockchains, as pseudonymity allows one individual or institution to operate multiple validators or miners. We do not account for potential Sybil attacks, as no scalable anti-Sybil systems exist for blockchains.\footnote{See Vitalik Buterin's blogs on \href{https://vitalik.eth.limo/general/2019/11/22/progress.html}{hard problems in cryptocurrency} and \href{https://vitalik.eth.limo/general/2023/07/24/biometric.html}{proof of personhood}.} Our measures of decentralization are thus best-case approximations. However, substantial Sybil-ing is improbable for the large blockchains we study, as participation requires costly resources (e.g., for PoS blockchains, Sybil costs scale with token market capitalization, which is in the trillions of dollars).\footnote{51\% attacks have occurred in the past for some other blockchains. See an article for Ethereum Classic's 51\% attack on August, 2020, on \href{https://www.coindesk.com/markets/2020/08/07/ethereum-classic-attacker-successfully-double-spends-168m-in-second-attack-report/}{coindesk.com} and a report for Bitcoin Gold on January, 2020, on \href{https://gist.github.com/metalicjames/71321570a105940529e709651d0a9765}{github.com}. Accessed April 22, 2024.} Moreover, our estimations of treatment effects are unlikely to be affected since Sybil vulnerabilities should be the same across pre- and post-treatment periods and across treatment and control blockchains.

Regarding generalizability, our findings are derived from a small number of blockchains observed during specific shock events. While these blockchains represent the largest and most economically significant consensus networks, the contextual idiosyncrasies of each quasi-experiment (a national policy ban, a single infrastructure provider's decision, and a long-planned protocol upgrade) limit the extent to which our results can be extrapolated to all blockchain systems or shock types. We view our findings as strongly suggestive evidence for the role of resource flexibility, with external validity to be established by future research across a broader set of blockchains and disruptions.

Beyond these primary limitations, our study focuses on consensus mechanisms and does not address other aspects of blockchain architecture such as smart contract design, interoperability, or scalability \citep{sai2021taxonomy, cong2022scaling}. While the consensus layer is structurally agnostic to the application layer, forward-looking participants may have partially factored in application-layer differences between Bitcoin and Ethereum when making relocation decisions. Future research should assess the long-term implications of design choices on blockchain decentralization and determine whether the observed benefits of flexible consensus mechanisms remain consistent over extended periods and across a broader set of blockchain ecosystems.

\subsection{Conclusion}

As blockchains increasingly underpin critical financial infrastructure, the ability to sustain decentralization under adversity is not merely a technical concern but an economic imperative. Our study provides strongly suggestive evidence that resource flexibility, a design-level property of consensus mechanisms, shapes how blockchains respond to and recover from exogenous shocks. Across three independent disruptions, we observe a consistent gradient: systems whose consensus resources have lower asset specificity recover faster and maintain broader participation. At the same time, resource flexibility alone does not guarantee redistribution of power among top validators, pointing to the need for complementary mechanisms. We hope this work encourages blockchain designers, operators, and policymakers to consider the flexibility of consensus resources as a first-order design parameter for sustaining decentralization.

\clearpage
\begingroup
\setstretch{1.0}

\endgroup

\clearpage
\appendix
\setcounter{page}{1}

\section*{\Large Appendix}

\large for \textit{Explaining Sustained Blockchain Decentralization with Quasi-Experiments: The Resource Flexibility of Consensus Mechanisms}\\

\vspace{1em}

\addcontentsline{toc}{section}{Appendix}
\renewcommand{\thesubsection}{\Alph{subsection}}
\renewcommand\thefigure{\thesubsection\arabic{figure}}
\renewcommand\thetable{\thesubsection\arabic{table}}
\renewcommand\theequation{\thesubsection\arabic{equation}}
\setcounter{figure}{0}
\setcounter{table}{0}
\setcounter{equation}{0}

\subsection{Literature Overview} \label{sec_appendix_literature}
\setcounter{table}{0}

Table~\ref{tb_literature} summarizes key studies on blockchain decentralization, their methodological approaches, main findings, and boundary conditions.

\begin{table}[ht]
\centering
\footnotesize
\begin{threeparttable}
\begin{tabular}{p{3.2cm}p{3cm}p{5cm}p{3.5cm}}
\hline
\hline \\[-1.8ex]
\textit{Study} & \textit{Approach} & \textit{Key Findings} & \textit{Boundary Conditions} \\
\hline \\[-1.8ex]
\textbf{\citet{gencer2018decentralization}}  & Empirical measurement & Bitcoin and Ethereum networks are not as decentralized as assumed; mining pools dominate & Snapshot analysis; limited time period \\
\textbf{\citet{cong2021decentralized}}  & Theoretical model & Decentralized consensus can emerge despite economies of scale through appropriate mechanism design & Model assumptions; permissionless setting \\
\textbf{\citet{cong2022scaling}}  & Causal identification & Layer-2 scaling increases decentralization in oracle data providers & Oracle subsystem only; not consensus layer \\
\textbf{\citet{capponi2023pow}}  & Theoretical model & Mining technology (ASICs vs. GPUs) affects decentralization through economies of scale & PoW blockchains only \\
\textbf{\citet{garratt2023fixed}}  & Theoretical model & Fixed costs in mining increase resilience to 51\% attacks when reward prices drop & Shocks to rewards, not resources \\
\textbf{\citet{mueller2024understanding}}  & Agent-based simulation & Higher participation promotes decentralization in PoS; wealth concentration is a risk & Simulation; PoS only \\
\textbf{\citet{chen2021governance}}  & Empirical (blockchain platforms) & Inverted U-shaped relationship: semi-decentralization outperforms full decentralization & Platform governance, not consensus \\
\textbf{\citet{sai2021taxonomy}}  & Taxonomy/Framework & Identifies multiple dimensions of decentralization across blockchain subsystems & Descriptive; no causal claims \\
\textbf{\citet{ju2025decentralizing}}  & Longitudinal empirical & Crypto ecosystems show mixed trends; recent centralization in consensus, NFTs, developers & Measurement focus; limited causal identification \\
\textbf{\citet{halaburda2020multidimensional}}  & Conceptual commentary & Blockchain decentralization is multidimensional; centralization preferable for system development, decentralization for transaction validation & Conceptual; no empirical analysis \\
\textbf{\citet{hsieh2023future}}  & Mixed methods (fsQCA) & Different dimensions of decentralization differently affect early-stage platform growth & Early-stage platforms; not consensus layer \\
\textbf{This paper}  & Quasi-experimental & Resource flexibility of consensus mechanisms enables sustained decentralization & Three specific shocks; limited blockchains \\
\hline
\hline
\end{tabular}
\begin{tablenotes}
\small
\item Notes: This table summarizes key studies on blockchain decentralization, their methodological approaches, main findings, and limitations.
\end{tablenotes}
\end{threeparttable}
\vspace{1em}
\caption{Overview of literature on blockchain decentralization} \label{tb_literature}
\end{table}

\subsection{Robustness Check with Consensus Layer Covariates} \label{sec_appendix_consensus_covar}

To test the robustness of our main difference-in-difference results of China's mining ban in Section~\ref{sec_results_china}, we include consensus layer covariates in our analysis. As explained in more detail in Section~\ref{sec_experimental_setting}, our analyses focus on the consensus layers of Bitcoin and Ethereum. While the two blockchains differ in the types of applications they support, the consensus layer is agnostic to the application layer. The consensus layer simply manages the secure production of blocks \citep{nakamoto2008bitcoin, buterin2014ethereum} and indirectly the economic incentives of block production.

In our robustness tests, we include pre-ban z-scored averages of daily metrics related to block production and economic incentives as covariates to isolate the effect of resource flexibility in response to China's ban. For block production, metrics include the number of transactions, the amount of data stored in megabytes (MB), and the estimated hashrates in megahashes per second (MH/s).\footnote{See Section~\ref{sec_data} for an explanation of hashrates. For reference, the hashrates were 174 exahashes per second for Bitcoin and 632 terahashes per second for Ethereum at their peak before the ban.} For economic incentives, we use the average transaction fee in US dollars (\$), excluding token price and block reward due to severe multicollinearity ($VIFs > 167,000$).\footnote{We excluded block reward and token price, as miner revenue can be derived from the transaction fee and number of transactions, avoiding redundancy.} The Bitcoin metrics were obtained from \href{https://www.blockchain.com/explorer/charts/}{blockchain.com}, and the Ethereum metrics were obtained from \href{https://etherscan.io/charts}{etherscan.io}.\footnote{Only weekly metrics were available for Bitcoin, so the daily metrics were forward-filled from the weekly metrics.} We remove the Exposure variable, as it is directly derived from hashrates and captured by our covariates.

In our robustness check, we find that our overall results are robust to these covariates (Table~\ref{tb_consensus_covar}) with and without interactions with the treatment variable (Table~\ref{tb_consensus_covar_interact}). The treatment effects on entropy, the Nakamoto coefficient, and HHI are all qualitatively similar but reduced when interactions are added. Interestingly, the treatment effects disappear for the $\text{Nodes}$ and $\text{Gini}$ metrics. This attenuation is consistent with these metrics being partially driven by consensus-layer characteristics that correlate with both the treatment and participation patterns. Our primary measure, Shannon entropy, remains robust to covariate interactions, reinforcing its use as our preferred decentralization metric. This result suggests two nuanced effects and highlights the importance of using multiple decentralization metrics to understand decentralization. First, the shock had an indistinguishable effect on the number of nodes in both Bitcoin and Ethereum. Second, the inequality in block production among the remaining nodes, as measured by the Gini coefficient, remains the same. 

\begin{table}[ht]
\centering
\resizebox{\textwidth}{!}{
\begin{threeparttable}
\begin{tabular}{lcccccccccc}
\hline
\hline \\[-1.8ex]
\textit{Dependent variable} & \multicolumn{2}{c}{\textit{Entropy}} & \multicolumn{2}{c}{\textit{Nodes}} & \multicolumn{2}{c}{\textit{Gini}} & \multicolumn{2}{c}{\textit{Nakamoto}} & \multicolumn{2}{c}{\textit{HHI}} \\
                                    & (1)  & (2) & (3) & (4) & (5) & (6) & (7) & (8) & (9) & (10) \\
\hline \\[-1.8ex]
\textbf{Treatment}  & -0.209$^{***}$ & -0.152$^{**}$ & 5.050$^{***}$ & 0.134 & 0.047$^{***}$ & 0.005 & -0.414$^{***}$ & -0.477$^{***}$ & 0.012$^{***}$ & 0.010$^{*}$\\
 & (0.049) & (0.048) & (1.229) & (1.112) & (0.010) & (0.010) & (0.094) & (0.117) & (0.003) & (0.004)\\
\textbf{Bitcoin}  & 0.020 & 0.296$^{**}$ & -33.980$^{***}$ & -33.920$^{***}$ & -0.327$^{***}$ & -0.370$^{***}$ & 1.577$^{***}$ & 1.776$^{***}$ & -0.037$^{***}$ & -0.055$^{***}$\\
 & (0.018) & (0.103) & (1.091) & (1.951) & (0.009) & (0.025) & (0.053) & (0.298) & (0.002) & (0.007)\\
\textbf{After}  & 0.020 & -0.045 & -6.893$^{***}$ & -3.501$^{**}$ & -0.033$^{***}$ & -0.011 & -0.060 & 0.089 & 0.001 & 0.004\\
 & (0.020) & (0.028) & (1.061) & (1.200) & (0.006) & (0.008) & (0.061) & (0.059) & (0.003) & (0.003)\\
\textbf{Intercept}  & 3.722$^{***}$ & 3.672$^{***}$ & 53.740$^{***}$ & 60.209$^{***}$ & 0.801$^{***}$ & 0.848$^{***}$ & 2.990$^{***}$ & 2.902$^{***}$ & 0.131$^{***}$ & 0.140$^{***}$\\
 & (0.015) & (0.069) & (0.948) & (1.627) & (0.005) & (0.013) & (0.008) & (0.249) & (0.002) & (0.007)\\
\textbf{Block Size (MB)}  &  & 0.147 &  & 1.167 &  & 0.002 &  & 0.467$^{*}$ &  & -0.010\\
 &  & (0.075) &  & (1.148) &  & (0.015) &  & (0.212) &  & (0.005)\\
\textbf{Hashrate (MH/s)}  &  & -0.004$^{***}$ &  & -0.034$^{***}$ &  & 0.000 &  & -0.006$^{***}$ &  & 0.000$^{***}$\\
 &  & (0.000) &  & (0.007) &  & (0.000) &  & (0.001) &  & (0.000)\\
\textbf{Number of Tx}  &  & 0.000 &  & -0.000$^{**}$ &  & -0.000$^{**}$ &  & 0.000 &  & -0.000\\
 &  & (0.000) &  & (0.000) &  & (0.000) &  & (0.000) &  & (0.000)\\
\textbf{Price (\$)}  &  & 0.000 &  & 0.000$^{***}$ &  & 0.000$^{***}$ &  & 0.000 &  & -0.000\\
 &  & (0.000) &  & (0.000) &  & (0.000) &  & (0.000) &  & (0.000)\\
\textbf{Fee per Tx (\$)}  &  & -0.000 &  & -0.002 &  & -0.000 &  & -0.003 &  & 0.000$^{*}$\\
 &  & (0.001) &  & (0.020) &  & (0.000) &  & (0.003) &  & (0.000)\\
\textbf{Reward (\$)}  &  & 0.000 &  & -0.000$^{**}$ &  & -0.000$^{**}$ &  & -0.000 &  & -0.000\\
 &  & (0.000) &  & (0.000) &  & (0.000) &  & (0.000) &  & (0.000)\\
\hline \\[-1.8ex]
Observations                & 1202 & 1202 & 1202 & 1202 & 1202 & 1202 & 1202 & 1202 & 1202 & 1202 \\
\hline
\hline
\end{tabular}
\begin{tablenotes}
\small
\item Notes: $^{*}$p$<$0.05; $^{**}$p$<$0.01; $^{***}$p$<$0.001. Standard errors are clustered by blockchain and month.
\end{tablenotes}
\end{threeparttable}
}
\vspace{1em}
\caption{Robustness check of difference-in-difference estimation of China's mining ban in Section~\ref{sec_results_china} with consensus layer covariates.} \label{tb_consensus_covar}
\end{table}

\begin{table}[ht]
\centering
\resizebox{\textwidth}{!}{
\begin{threeparttable}
\begin{tabular}{lcccccccccc}
\hline
\hline \\[-1.8ex]
\textit{Dependent variable} & \multicolumn{2}{c}{\textit{Entropy}} & \multicolumn{2}{c}{\textit{Nodes}} & \multicolumn{2}{c}{\textit{Gini}} & \multicolumn{2}{c}{\textit{Nakamoto}} & \multicolumn{2}{c}{\textit{HHI}} \\
                                    & (1)  & (2) & (3) & (4) & (5) & (6) & (7) & (8) & (9) & (10) \\
\hline \\[-1.8ex]
\textbf{After}  & -0.118$^{***}$ & -0.118$^{***}$ & -1.673 & -1.673 & -0.009 & -0.009 & -0.105 & -0.105 & 0.013$^{**}$ & 0.013$^{**}$\\ 
 & (0.030) & (0.030) & (1.727) & (1.730) & (0.014) & (0.014) & (0.101) & (0.101) & (0.004) & (0.004)\\
\textbf{Bitcoin}  & 0.020 & 0.020 & -33.980$^{***}$ & -33.980$^{***}$ & -0.327$^{***}$ & -0.327$^{***}$ & 1.577$^{***}$ & 1.577$^{***}$ & -0.037$^{***}$ & -0.037$^{***}$\\ 
 & (0.017) & (0.017) & (1.019) & (1.021) & (0.007) & (0.007) & (0.051) & (0.051) & (0.002) & (0.002)\\
\textbf{Intercept}  & 3.722$^{***}$ & 3.722$^{***}$ & 53.740$^{***}$ & 53.740$^{***}$ & 0.801$^{***}$ & 0.801$^{***}$ & 2.990$^{***}$ & 2.990$^{***}$ & 0.131$^{***}$ & 0.131$^{***}$\\ 
 & (0.011) & (0.011) & (0.908) & (0.909) & (0.005) & (0.005) & (0.016) & (0.016) & (0.001) & (0.001)\\
\textbf{Treatment}  & -0.109 & -0.022 & -0.989 & -0.203 & -0.001 & -0.000 & -0.253 & -0.052 & 0.002 & 0.000\\ 
 & (0.063) & (0.013) & (1.554) & (0.319) & (0.013) & (0.003) & (0.145) & (0.030) & (0.005) & (0.001)\\
\textbf{Treatment $\times$ Block Size}  &  & -0.008 &  & -0.070 &  & -0.000 &  & -0.018 &  & 0.000\\ 
 &  & (0.004) &  & (0.110) &  & (0.001) &  & (0.010) &  & (0.000)\\
\textbf{Treatment $\times$ Fees per Tx}  &  & -0.005 &  & -0.049 &  & -0.000 &  & -0.013 &  & 0.000\\ 
 &  & (0.003) &  & (0.077) &  & (0.001) &  & (0.007) &  & (0.000)\\
\textbf{Treatment $\times$ Hash Rate}  &  & -0.039 &  & -0.357 &  & -0.000 &  & -0.091 &  & 0.001\\ 
 &  & (0.023) &  & (0.561) &  & (0.005) &  & (0.052) &  & (0.002)\\
\textbf{Treatment $\times$ \# Txs}  &  & 0.017 &  & 0.157 &  & 0.000 &  & 0.040 &  & -0.000\\ 
 &  & (0.010) &  & (0.248) &  & (0.002) &  & (0.023) &  & (0.001)\\
\textbf{Block Size}  & -0.002 & -0.002 & -0.259 & -0.259 & -0.001 & -0.001 & 0.017 & 0.017 & 0.000 & 0.000\\ 
 & (0.009) & (0.009) & (0.240) & (0.240) & (0.003) & (0.003) & (0.046) & (0.047) & (0.001) & (0.001)\\
\textbf{Fees per Tx}  & 0.007 & 0.007 & -0.849$^{*}$ & -0.849$^{*}$ & -0.003 & -0.003 & -0.045 & -0.045 & -0.000 & -0.000\\ 
 & (0.008) & (0.008) & (0.425) & (0.425) & (0.003) & (0.003) & (0.023) & (0.023) & (0.001) & (0.001)\\
\textbf{Hash Rate}  & 0.030$^{**}$ & 0.030$^{**}$ & 0.097 & 0.097 & 0.008 & 0.008 & -0.027 & -0.027 & -0.002 & -0.002\\ 
 & (0.010) & (0.010) & (0.443) & (0.444) & (0.006) & (0.006) & (0.036) & (0.037) & (0.001) & (0.001)\\
\textbf{\# Txs}  & 0.022$^{*}$ & 0.022$^{*}$ & -1.212$^{***}$ & -1.212$^{***}$ & -0.015$^{***}$ & -0.015$^{***}$ & 0.081$^{*}$ & 0.081$^{*}$ & -0.003$^{**}$ & -0.003$^{**}$\\ 
 & (0.008) & (0.008) & (0.281) & (0.281) & (0.003) & (0.003) & (0.032) & (0.032) & (0.001) & (0.001)\\
\hline \\[-1.8ex]
$R^2$ & 0.306 & 0.306 & 0.962 & 0.962 & 0.953 & 0.953 & 0.727 & 0.727 & 0.687 & 0.687 \\
Observations & 1202 & 1202 & 1202 & 1202 & 1202 & 1202 & 1202 & 1202 & 1202 & 1202 \\
\hline
\hline
\end{tabular}
\begin{tablenotes}
\small
\item Notes: $^{*}$p$<$0.05; $^{**}$p$<$0.01; $^{***}$p$<$0.001. Standard errors are clustered by blockchain and month. All consensus layer covariates are pre-ban averages, normalized as a Z-score for each blockchain. \textit{Tx} refers to transactions. Block size is in megabytes, and hashrate is in megahashes per second.
\end{tablenotes}
\end{threeparttable}
}
\vspace{1em}
\caption{Robustness check of difference-in-difference estimation of China's mining ban in Section~\ref{sec_results_china} with consensus layer covariates and covariate-treatment interactions.}
\label{tb_consensus_covar_interact}
\end{table}

\setcounter{figure}{0} 
\setcounter{table}{0} 
\setcounter{equation}{0} 
\clearpage

\subsection{Identifying Ethereum Validators post Proposer-Builder Separation (PBS)} \label{sec_appendix_mev}

Proposer-builder separation (PBS) was designed to create an open marketplace for block building. Critically, block building involves the ordering of transactions, which can extract arbitrary opportunities by front-running, sandwiching, or other methods. This value extraction through transaction ordering is called maximal extractable value (MEV). Flashbots, who first studied MEV on Ethereum \citep{daian2019mev}, developed an initial implementation of  PBS in a middleware called MEV-Boost.\footnote{See the GitHub \href{https://github.com/flashbots/mev-boost}{repository} for MEV-Boost. Accessed April 25, 2024.} MEV-Boost was deployed on Ethereum on the same day as the Merge on block 15537940.\footnote{See deployment \href{https://etherscan.io/block/0x7d57a1d26f71724737f5dc780ca2dfb778c2fc5be29bcaeb7b989b768953aabe}{transaction} on Etherscan. See also the \href{https://boost.flashbots.net/mev-boost-status-updates/mev-boost-status-update-sep-9-sept-22-2022}{blog post} by Flashbots. Accessed April 25, 2024.} By running MEV-Boost, validators can access a competitive block-building market and sell block space to arbitragers to increase staking rewards by over 60\%.\footnote{See article by Flashbots on \href{https://hackmd.io/@flashbots/mev-in-eth2}{hackmd.io}. Accessed April 25, 2024.}

MEV-Boost creates problems for data collection of validators because the block reward recipient now becomes the block builder, not the proposer (also known as the validator). However, in the context of our study on the consensus layer, we need to identify the proposer for each block, not the builder. In the block, the builder first receives the block reward, and the builder transfers the block reward plus additional fees for the proposer as a transaction in the block.\footnote{Because the proposer ultimately signs the block, the proposer is guaranteed to receive the block reward in the same block that it validates by checking whether it receives the reward before signing the block.} Thus, for a block that has an MEV builder as the block recipient, we identify the proposer within the block transactions in the following way.

First, we obtain a list of labeled addresses for MEV builders on Etherscan.\footnote{See \href{https://etherscan.io/accounts/label/mev-builder}{etherscan.io}. Accessed April 25, 2024} The list of MEV builders that we use is available at \href{https://dune.com/queries/3665816}{https://dune.com/queries/3665816}. Because builders are in a competitive marketplace, they are incentivized to reveal their identities as builders.\footnote{While Ethereum allows anonymous or private builders, such builders should be rare, if they exist at all.} Then, we obtain all blocks whose recipients are \textit{not} on the list of MEV builders. These blocks were not sold to builders and have validators as the block reward recipients.

Then, we obtain a list of all blocks whose block reward recipients are in the list of MEV builders. These blocks \textit{were} sold to builders and have the builder as a block reward recipient. In this case, we identified the transaction by which the builder sent the block reward to the proposer in a few different ways. First, if there was a transaction in an MEV block whose sender is a builder and the receiver is a known proposer (see below for how we obtain a list of proposers), then we label the recipient of the transaction as the proposer for that block. Second, a few builders sent the block reward from another address that it owned, presumably to save money on transaction fees. This was common for larger MEV builders. We manually identify the separate addresses for MEV builders, which are available at \href{https://dune.com/queries/3669067}{https://dune.com/queries/3669067}. Lastly, some MEV builders are also proposers. For these builders, we label the builder as also the proposer of the block; such builders are labeled at \href{https://dune.com/queries/3665820}{https://dune.com/queries/3665820}. These methods accounted for all of the blocks that we obtain from January 1, 2022, to March 31, 2023. The SQL implementation of these methods can be found at \href{https://dune.com/queries/3664358}{https://dune.com/queries/3664358}.

As mentioned above, we obtain a list of block reward recipients in the following ways. First, we query all addresses that deposited ETH to the Beacon staking contract on Ethereum. Now, if Ethereum did not have staking pools or smart contracts, then this list would be sufficient. Second, we obtain a list of the top proposer fee recipients from Etherscan, which may include staking pools.\footnote{See list on \href{https://etherscan.io/accounts/label/proposer-fee-recipient}{etherscan.io}. Accessed April 25, 2024.} Most, if not all, staking pools are known because they have the incentive to obtain as much stake in their pool as possible and thus market themselves to potential stakers. Third, we have manually labeled a list of MEV builders who are also recipients of block rewards. They are clearly recipients because they have received numerous fees from other labeled MEV builders. See address \href{https://etherscan.io/address/0x7e2a2FA2a064F693f0a55C5639476d913Ff12D05}{0x7e2a2FA2a064F693f0a55C5639476d913Ff12D05} as an example. The final list of recipients is available at \href{https://dune.com/queries/3665820}{https://dune.com/queries/3665820}.

\setcounter{figure}{0} 
\setcounter{table}{0} 
\setcounter{equation}{0} 
\clearpage

\subsection{Post-PBS Analysis of MEV Builders} \label{sec_appendix_postpbs}

\begin{figure}[!ht]
\centering
{\includegraphics[width=\linewidth]{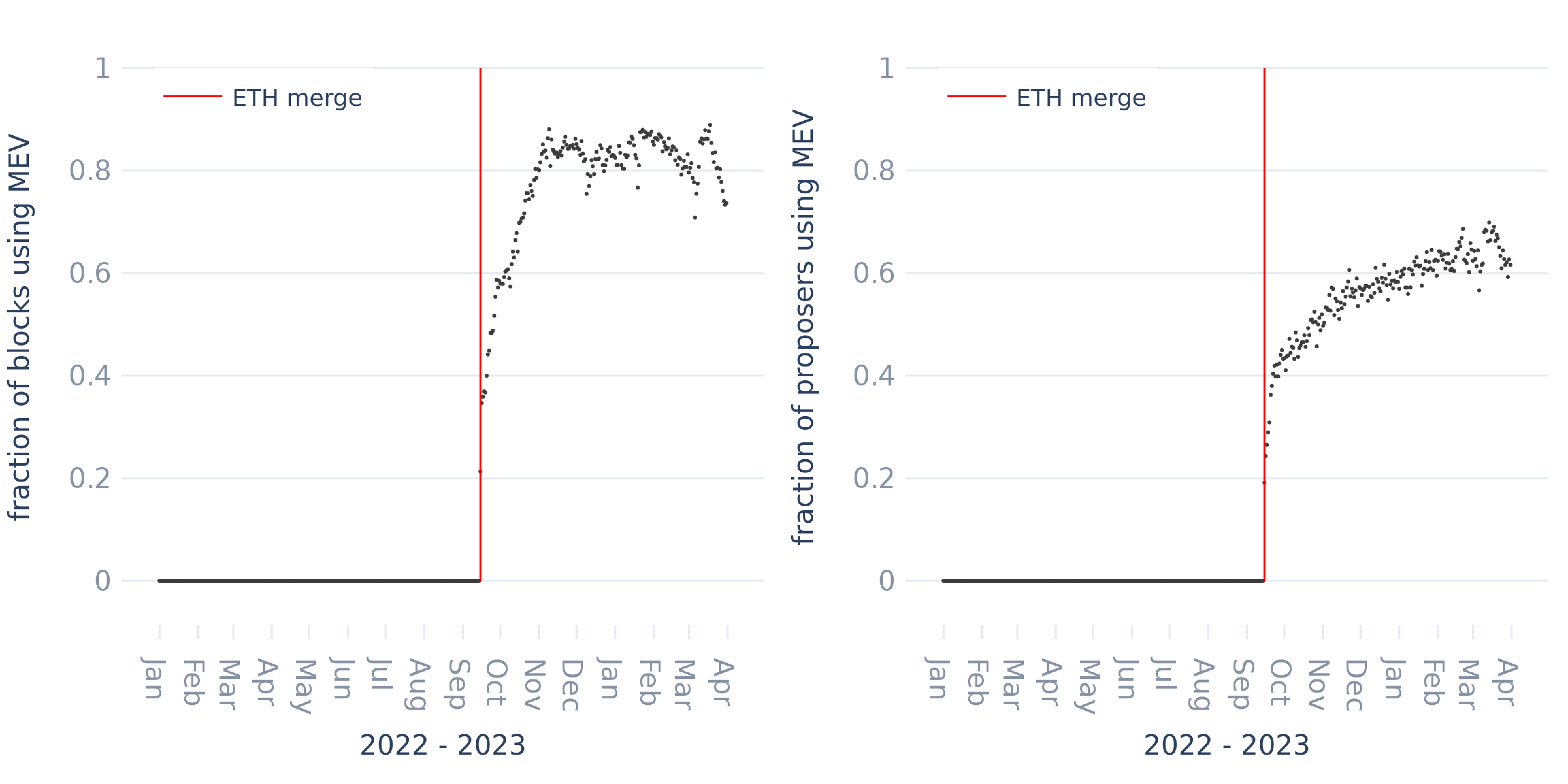}}
\caption{\small (Left) Fraction of blocks using MEV. (Right) fraction of proposers using MEV.} 
\label{fig_mev_blocks}
\end{figure}

As explained in Section~\ref{sec_appendix_mev}, proposer-builder separation (PBS) was designed to create a competitive marketplace for block builders to ameliorate the negative externalities of maximal extractable value (MEV). Here, we report data on the MEV builders involved after the implementation of MEV-Boost, which is coincident on the same day as the Ethereum Merge. Figure~\ref{fig_mev_blocks} shows the daily fraction of blocks that have used MEV (left) and the daily fraction of proposers (\textit{i.e.}, validators) that have produced at least one MEV block.

\begin{figure}[!ht]
\centering
{\includegraphics[width=\linewidth]{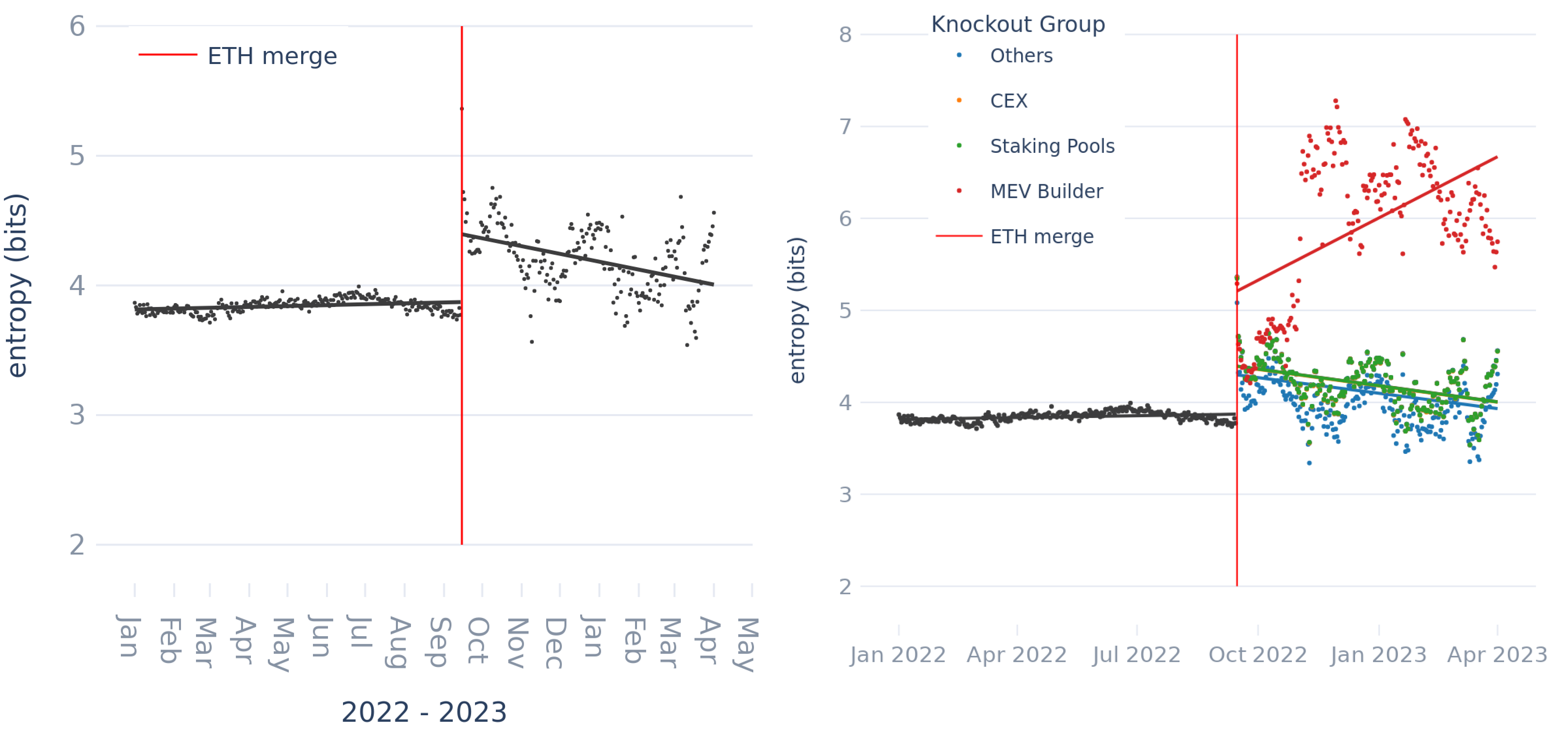}}
\caption{\small Knockout simulations of Ethereum's entropy after the Merge.} 
\label{fig_merge_knockout}
\end{figure}

Then, to test the robustness of our analyses in Section~\ref{sec_results_merge}, we replicate the same analyses without correcting for PBS. In Figure~\ref{fig_merge_knockout}, we can see that entropy increases only by around 0.5 at the Merge. There is also a sharp decline in the entropy that corresponds with the increasing usage of MEV shown in Figure~\ref{fig_mev_blocks}.

For months following the Merge, there is also a steady, daily decrease in Ethereum's entropy of $-0.002^{***}$. To understand the small but statistically significant negative slope, we investigated whether any particular group of validators had a role in the negative slope of entropy after the Merge. To do so, we simulate ``knockout'' experiments in which we remove particular groups of validators from the analysis after the Merge and recompute the entropy. The groups included centralized exchanges (CEX), Staking Pools such as Lido, MEV (maximal extractable value) builders, and ``Others'' which include individuals or are unlabeled.\footnote{See \href{https://ethereum.org/en/developers/docs/mev/}{ethereum.org} for a detailed explanation of MEV. List of sequencers from \href{https://etherscan.io/accounts/label/mev-builder?subcatid=undefined&size=100&start=0&col=1&order=asc}{etherscan.io}. Accessed April 23, 2024.} We see in Figure~\ref{fig_merge_knockout} that removing MEV builders from the data had a substantial positive effect on both the slope and the absolute value of entropy. This result suggests that MEV builders are a \textit{centralizing} force on Ethereum's consensus. Removing CEX and Staking Pools validators has negligible effects while removing ``Others'' has a negative effect, as one can expect from removing individuals from consensus.

The results here not only validate our MEV analysis but also suggest that the composability of protocols through proposer-builder separation (PBS) or otherwise, while introducing opportunities for open protocols, market competition, and interoperability, opens up the risk of centralization in each component. This unexpected discovery, which emerged from our analysis, does not directly implicate the resource flexibility hypothesis but is an interesting element of the decentralization of blockchains that warrants further investigation. While such analysis is beyond the scope of this paper, we encourage such work in future research.

\setcounter{figure}{0} 
\setcounter{table}{0} 
\setcounter{equation}{0} 
\clearpage

\subsection{Multi-Period Difference-in-Difference} \label{sec_appendix_multiperiod}

To account for the rolling enforcement of China's policy change banning crypto mining, we decomposed our difference-in-difference estimation in Section~\ref{sec_results_china} into multiple periods. Instead of simply having a single date that divides the time series into $\text{After}_t=0$ before May 15, 2021, and $\text{After}_t=1$ after, we create both $\text{During}_t$ and $\text{After}_t$ variables. The $\text{During}_t$ variable is a binary indicator variable that equals one from May 15, 2021, to the end of July 2021 and represents the days that the ban was being enforced. The $\text{After}_t$ variable is a binary indicator variable as before, but it now equals one starting on August 2021. We also reformulated the $\text{Exposure}_{it}$ as the cross-term $(\text{During}_t \times \text{Exposure}_i)$, instead of $(\text{After}_t \times \text{Exposure}_i)$, to account for the exposure occurring during the rolling ban. We then specified a multi-period difference-in-difference estimation by
\begin{equation} \label{eq_did_mp}
\begin{split}
\text{Entropy}_{it} =&\; \delta_1 (\text{During}_t \times \text{Bitcoin}_i) + \delta_2 (\text{After}_t \times \text{Bitcoin}_i) \\
& + \beta_1 \text{Exposure}_{it} + \beta_2 \text{Bitcoin}_{i} + \beta_3 \text{During}_t + \beta_4 \text{After}_t + \epsilon_{it}
\text{,}
\end{split}
\end{equation}
where the variables are the same as explained in Equation~\ref{eq_did}.

When we estimate Equation~\ref{eq_did_mp}, we find a similar result as in Section~\ref{sec_results_china} of the main manuscript. In Column 1 of Table~\ref{tb_china_multiperiod}, we observe that there were no significant effects during the rollout, as seen in the following terms: $\text{During}$, $(\text{During} \times \text{Bitcoin})$, and $(\text{During} \times \text{Exposure})$. However, we did see a negative effect of $(\text{After} \times \text{Bitcoin})$ after the rollout. This negative effect is similar to our results in Section~\ref{sec_results_china} but also slightly larger because it now excludes the more minor effects during the rollout of the ban. The consistency of the negative effect observed post-rollout with our earlier findings in Section~\ref{sec_results_china} not only validates our initial results but also suggests that the impact of the ban on Bitcoin's decentralization was more pronounced when isolating the post-rollout period, thereby reinforcing the robustness of our analysis.

\begin{table}[ht]
\centering
\resizebox{\textwidth}{!}{
\begin{threeparttable}
\begin{tabular}{lcccccccccc}
\hline
\hline \\[-1.8ex]
\textit{Dependent variable} & \multicolumn{2}{c}{\textit{Entropy}} & \multicolumn{2}{c}{\textit{Nodes}} & \multicolumn{2}{c}{\textit{Gini}} & \multicolumn{2}{c}{\textit{Nakamoto}} & \multicolumn{2}{c}{\textit{HHI}} \\
                                    & (1)  & (2) & (3) & (4) & (5) & (6) & (7) & (8) & (9) & (10) \\
\hline \\[-1.8ex]
\textbf{After}  & 0.078$^{*}$ & -0.223$^{***}$ & -5.800$^{***}$ & -3.774 & -0.041$^{***}$ & 0.003 & -0.050 & -0.274 & 0.001 & 0.029$^{**}$\\ 
 & (0.033) & (0.057) & (1.411) & (3.737) & (0.007) & (0.016) & (0.084) & (0.307) & (0.004) & (0.009)\\
\textbf{Treatment (After)}  & -0.497$^{***}$ & 0.180 & 1.219 & 2.251 & 0.047$^{***}$ & -0.013 & -0.832$^{***}$ & 0.104 & 0.027$^{***}$ & -0.028$^{*}$\\ 
 & (0.063) & (0.143) & (1.575) & (4.272) & (0.011) & (0.030) & (0.128) & (0.350) & (0.005) & (0.012)\\
\textbf{Bitcoin}  & 0.020 & -0.221$^{***}$ & -33.980$^{***}$ & -34.347$^{***}$ & -0.327$^{***}$ & -0.306$^{***}$ & 1.577$^{***}$ & 1.244$^{***}$ & -0.037$^{***}$ & -0.017$^{***}$\\ 
 & (0.026) & (0.057) & (1.309) & (2.064) & (0.010) & (0.012) & (0.062) & (0.118) & (0.003) & (0.004)\\
\textbf{Day}  &  & 0.002$^{***}$ &  & -0.002 &  & -0.000$^{**}$ &  & 0.003$^{*}$ &  & -0.000$^{***}$\\ 
 &  & (0.000) &  & (0.016) &  & (0.000) &  & (0.001) &  & (0.000)\\
\textbf{During}  & 0.010 & -0.140$^{***}$ & -6.520$^{***}$ & -5.509$^{*}$ & -0.021$^{*}$ & 0.001 & 0.010 & -0.102 & -0.007$^{*}$ & 0.007\\ 
 & (0.027) & (0.033) & (1.612) & (2.627) & (0.008) & (0.011) & (0.010) & (0.115) & (0.003) & (0.005)\\
\textbf{Treatment (During)}  & -0.026 & 0.312$^{***}$ & 6.760$^{***}$ & 7.276$^{*}$ & 0.035$^{***}$ & 0.006 & -0.138$^{*}$ & 0.330$^{*}$ & 0.010$^{**}$ & -0.018$^{**}$\\ 
 & (0.039) & (0.079) & (1.669) & (2.825) & (0.011) & (0.015) & (0.066) & (0.159) & (0.004) & (0.006)\\
\textbf{Exposure $\times$ Day}  &  & -0.006$^{***}$ &  & -0.010 &  & 0.001$^{*}$ &  & -0.009$^{**}$ &  & 0.001$^{***}$\\ 
 &  & (0.001) &  & (0.034) &  & (0.000) &  & (0.003) &  & (0.000)\\
\textbf{Intercept}  & 3.722$^{***}$ & 3.828$^{***}$ & 53.740$^{***}$ & 53.020$^{***}$ & 0.801$^{***}$ & 0.785$^{***}$ & 2.990$^{***}$ & 3.070$^{***}$ & 0.131$^{***}$ & 0.121$^{***}$\\ 
 & (0.020) & (0.020) & (1.280) & (1.898) & (0.007) & (0.008) & (0.010) & (0.083) & (0.002) & (0.003)\\
\hline \\[-1.8ex]
Observations                & 1202 & 1202 & 1202 & 1202 & 1202 & 1202 & 1202 & 1202 & 1202 & 1202 \\
\hline
\hline
\end{tabular}
\begin{tablenotes}
\small
\item Notes: $^{*}$p$<$0.05; $^{**}$p$<$0.01; $^{***}$p$<$0.001. Standard errors are clustered by blockchain and month.
\end{tablenotes}
\end{threeparttable}
}
\vspace{1em}
\caption{China bans crypto mining} \label{tb_china_multiperiod}
\end{table}


While the above multi-period analysis does not reveal any significant effects during the rollout, this may be because the rollout is gradual. That is, this simple ``difference-in-means'' approach may not reveal the nuanced dynamics of the rollout, as we can visually see happening in Figure~\ref{fig_china}. Thus, to get an even more fine-grain understanding of the multi-period dynamics, we extended Equation~\ref{eq_did_mp} by adding the time variable $\text{Day}_t$, by itself and as cross-terms with $(\text{During}_t \times \text{Chain}_i)$ and $(\text{After}_t \times \text{Chain}_i)$. We also add the time variable as a cross-term with the control variable $\text{Exposure}_{it}$ because it is indeed the exposure that is being rolled out gradually during the ban. As such, we are even more explicitly estimating the effect of resource flexibility, or lack thereof, \textit{in response to} the exposure while, at the same time, even controlling for the time-varying effects of the exposure itself. We specified this new time-varying multi-period difference-in-difference estimation by
\begin{equation} \label{eq_did_mp_day}
\begin{split}
\text{Entropy}_{it} =&\;  \delta_1 (\text{During}_t \times \text{Bitcoin}_i \times \text{Day}_t) + \delta_2 (\text{After}_t \times \text{Bitcoin}_i \times \text{Day}_t) \\
& + \beta_1 (\text{Exposure}_{it} \times \text{Day}) +\beta_2 \text{Bitcoin}_{i} + \beta_3 \text{During}_t + \beta_4 \text{After}_t \\
& + \beta_5 \text{Day}_t + \epsilon_{it}
\text{.}
\end{split}
\end{equation}

By estimating Equation~\ref{eq_did_mp_day}, we first find that the exposure itself has the greatest negative effect of $-0.007$ bits per day. This initial result confirms the simple intuition that the direct shutdown of nodes should have a negative effect on decentralization. In fact, in this initial period, Bitcoin received this negative effect of the exposure slightly less than Ethereum did by $0.001$ bits per day ($p=0.045$), and this is visible in Figure~\ref{fig_china}, where we can see that despite Bitcoin's greater exposure, both Bitcoin and Ethereum experienced a similar decrease in entropy throughout the rollout period. However, after the rollout period, Bitcoin experienced a significant negative effect of $-0.002$ bits per day, consistent with our other results. Thus, although both Bitcoin and Ethereum initially saw a decline in decentralization, Bitcoin's more pronounced negative impact in the post-rollout period highlights the differing recovery dynamics and underscores the role of resource flexibility in influencing a blockchain's resilience to external shocks. Our decomposition analysis in Section~\ref{sec_appendix_decompose} of the Appendix reveals the behavior of individual nodes and mining pools that may further elucidate the effects of the ban even more precisely.

\setcounter{figure}{0} 
\setcounter{table}{0} 
\setcounter{equation}{0} 
\clearpage

\subsection{Bitcoin Hashrate Mining Maps} \label{sec_appendix_hashrate}

Here, we validate that the hashrates observed in Figure~\ref{fig_hashrate}, which we used to estimate the $\text{Exposure}$ covariate, do indeed represent exposure to China's mining ban. While we do not have geographical hashrates for Ethereum, we use the data available for Bitcoin as validation for our exposure measures using total hashrates. To validate our exposure measures, we use the \href{https://ccaf.io/cbnsi/cbeci/mining_map}{Bitcoin Mining Map} collected by The Cambridge Centre for Alternative Finance \citep{cbeci2023bitcoin}. The Centre has partnered with several Bitcoin mining pools and uses the IP (Internet Protocol) addresses of mining facility operators to identify the geographical location of a miner.

\begin{figure}[!ht]
\centering
{\includegraphics[width=\linewidth]{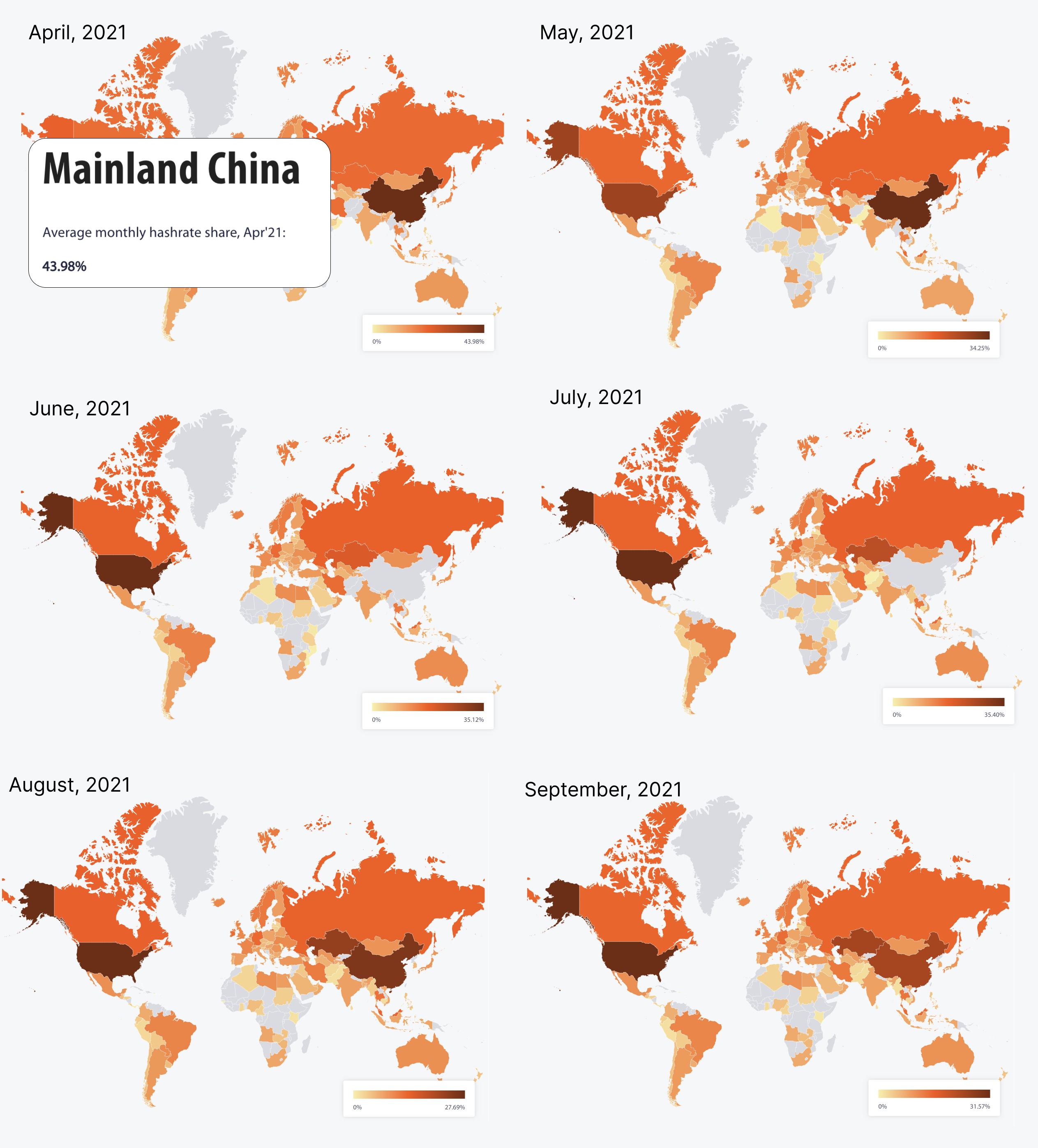}}
\caption{\small The Bitcoin Mining Map by The Cambridge Centre for Alternative Finance from April to September 2021.}
\label{fig_hashrate_map}
\end{figure}

\begin{figure}[!ht]
\centering
{\includegraphics[width=\linewidth]{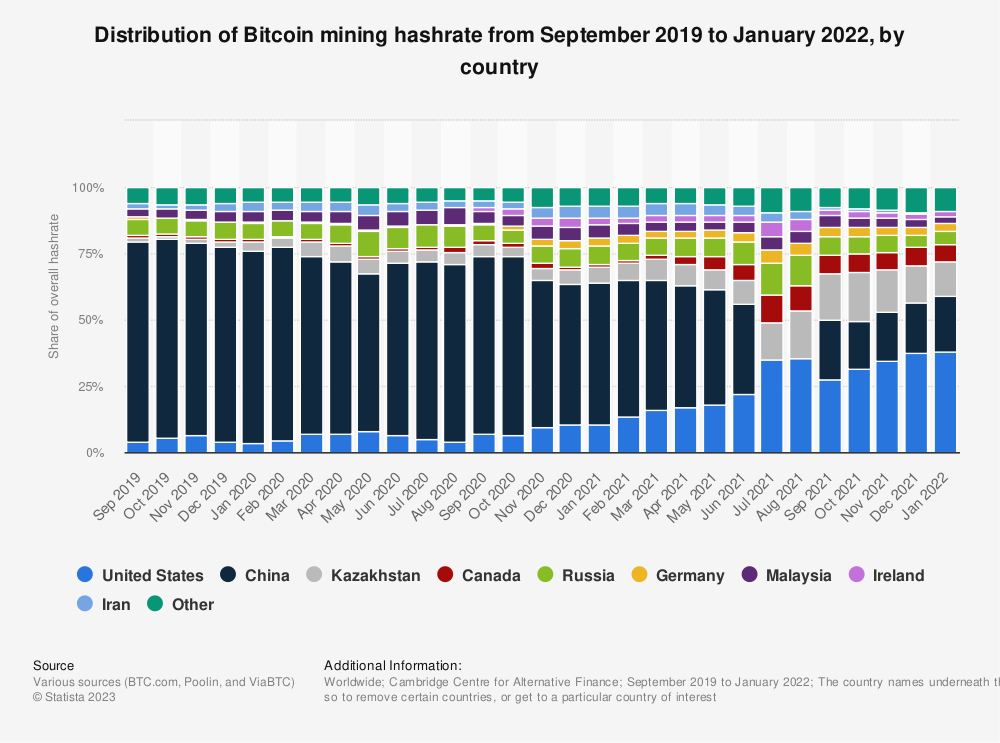}}
\caption{\small Distribution of Bitcoin mining hashrate from September 2019 to January 2022, by country.}
\label{fig_hashrate_country}
\end{figure}

Figure~\ref{fig_hashrate_map} shows the monthly hashrates by country from April 2021 to September 2021, around the time of China's mining ban. First, we observe on the map that monthly hashrates decreased from 43.98\% in April 2021 to 34.25\% in May 2021. Then, China's hashrate disappears to 0\% for the months of June and July until it recovers 22.29\% in August. As an alternative visualization, Figure~\ref{fig_hashrate_country} shows a chart from Statista that displays the same data as in Figure~\ref{fig_hashrate_map} as a bar chart \citep{cambridge2022distribution}. This trend accurately matches the decrease in total hashrates from May 2021 to July 2021 in Figure~\ref{fig_hashrate}. Second, the maximum percentage of China's hashrate was 43.98\%, which is close to the exposure measure of 51.14\% for Bitcoin. While it is conceivable that the partnered mining pools colluded with Chinese miners to hide their IP, that seems unlikely given the similarity between the hashrate of Chinese miners in the map and the total hashrate. The close alignment between the observed hashrate percentages in China and the total hashrate data further validates our exposure measure and our analysis surrounding the impact of China's mining ban on blockchain decentralization.

\setcounter{figure}{0} 
\setcounter{table}{0} 
\setcounter{equation}{0} 
\clearpage

\subsection{Decomposing China's Mining Ban} \label{sec_appendix_decompose}

To better understand how China's mining ban may have affected Bitcoin, we decompose the effects of the ban to the level of individual nodes. This analysis does not take the rolling nature of the ban into account and is certainly not causal. Indeed, because of Bitcoin's pseudonymity, it is difficult to determine the identity or geographic location of nodes. Moreover, mining pools can obfuscate the distribution of rewards to their participants and make it even more difficult to find definitive, causal observations at the node level. However, by exploring node-level data, we can check for the robustness of our interpretations of our results and also investigate more nuanced effects of the ban, such as those to mining pools versus individual miners.

\begin{figure}[!ht]
\centering
{\includegraphics[width=\linewidth]{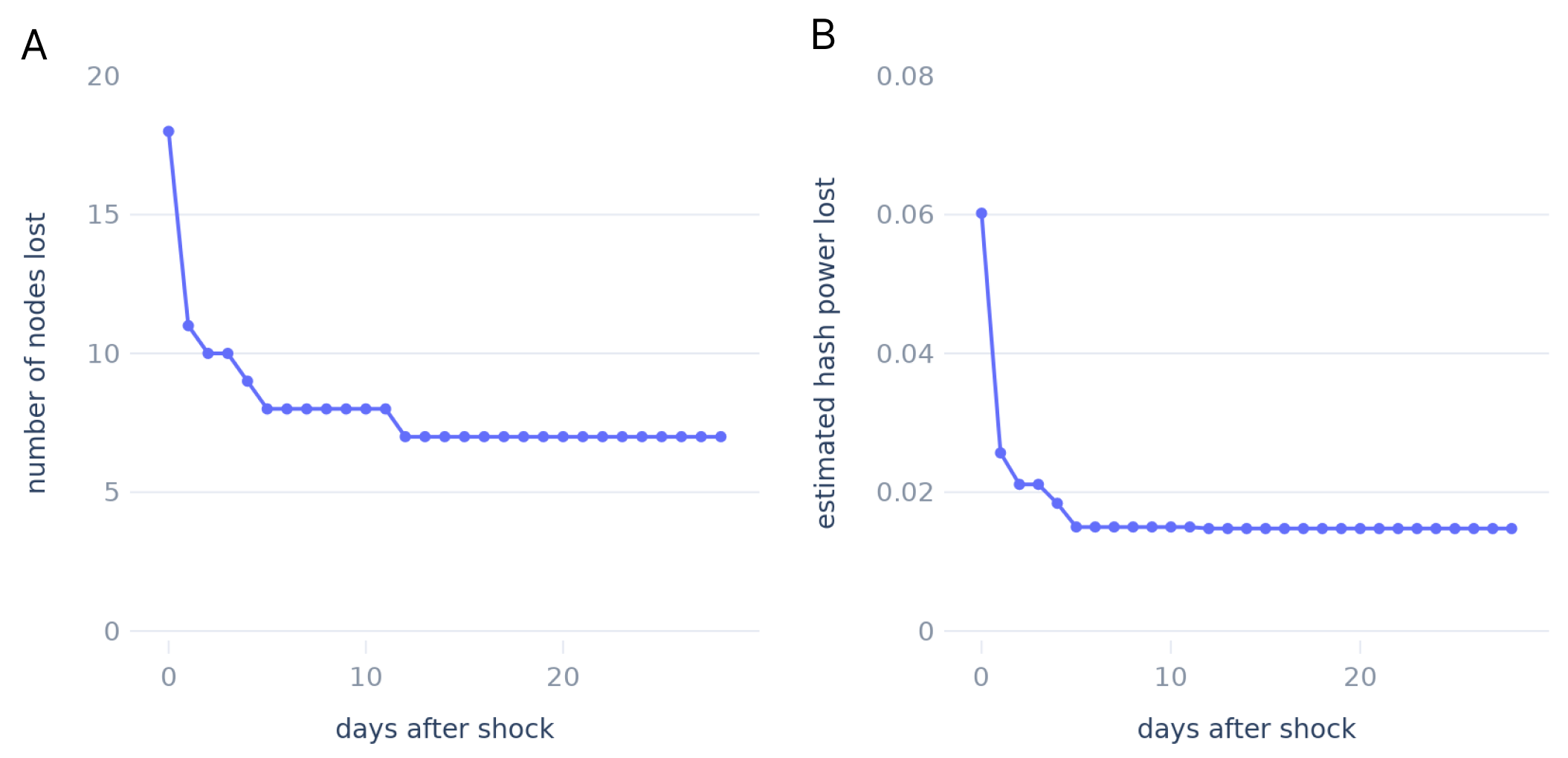}}
\caption{\small (A) The number of nodes lost after the initial shock. (B) Estimated hash power of the nodes lost after the initial shock.}
\label{fig_china_miners}
\end{figure}

First, we examine the nodes that may have individually been affected by the ban. To construct the number of nodes lost after the shock shown in Figure~\ref{fig_china_miners}, we determine the unique node addresses that received block rewards from 30 days prior to the shock on May 15, 2021, up to the day of the shock to see which nodes are no longer receiving any block rewards. If we conservatively assume that the initially high number of 18 nodes in Figure~\ref{fig_china_miners}A simply did not yet have a chance to receive rewards, we can see that there are around 7 nodes that no longer receive rewards at the steady state. Those nodes represent, however, only 0.0147 percent of the estimated hash power of the network: a far cry from the 51.1\% decrease in total hashrate we observe in Section~\ref{sec_variable}. 

\begin{figure}[!ht]
\centering
{\includegraphics[width=\linewidth]{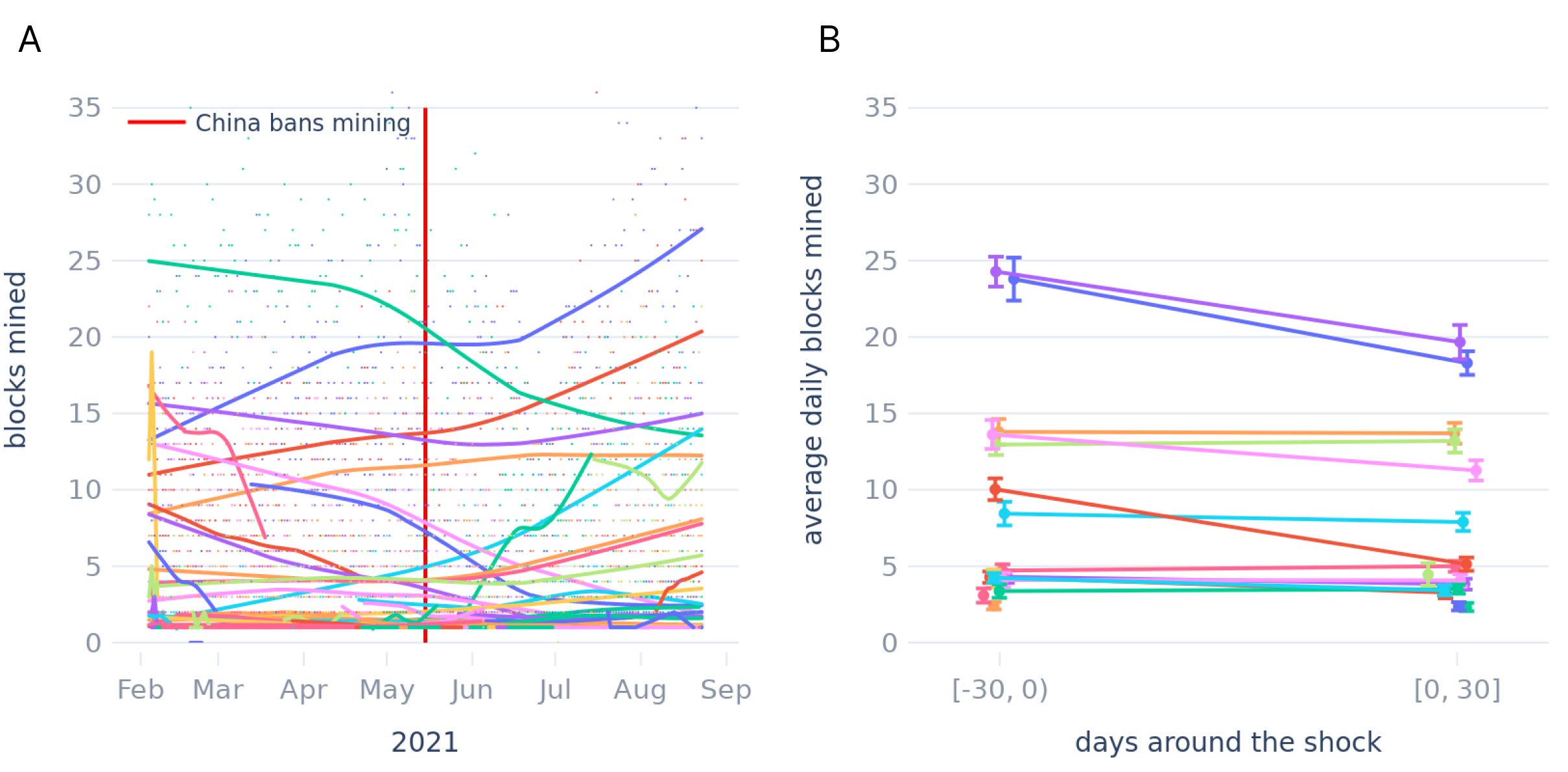}}
\caption{\small Number of blocks mined by Bitcoin miners around the time of the shock. (A) Daily number of blocks mined with Lowess trend lines. (B) The average number of blocks mined 30 days before and after the shock with standard errors.}
\label{fig_china_mining}
\end{figure}

Given that individual miners do not seem to be much affected by our initial estimation, we estimate whether mining pools may have experienced changes in their mining throughout the ban. In Figure~\ref{fig_china_mining}A, we display the daily number of blocks mined per node address around the time of the ban with lowess trend lines. Though we cannot make any causal claims, we can clearly observe that some nodes are producing more and some fewer blocks around the time of the ban. We also show summary statistics for 30 days prior and 30 days after the ban in Figure~\ref{fig_china_mining}B. These varying trends in block production around the time of the ban suggest that mining pools may have experienced differential impacts, underscoring the impact of the ban on mining pools versus individual miners.

While many individual nodes cannot be identified due to Bitcoin's pseudonymity, some larger nodes, such as mining pools, are labeled, allowing us to more accurately assess the impact of the ban on these specific entities within the blockchain network. We examine three examples of mining pools that have experienced a reduction in the daily nodes mined from our above analysis: the Binance pool, the Huobi pool, and an unknown pool.\footnote{The charts for these exact nodes can be found for the \href{https://www.blockchain.com/explorer/addresses/btc/1DSh7vX6ed2cgTeKPwufV5i4hSi4pp373h}{Binance}, \href{https://www.blockchain.com/explorer/addresses/btc/1EepjXgvWUoRyNvuLSAxjiqZ1QqKGDANLW}{Huobi}, and the \href{https://www.blockchain.com/explorer/addresses/btc/191sNkKTG8pzUsNgZYKo7DH2odg39XDAGo}{unlabeled} pools. Accessed October 30, 2023.} While we cannot definitively identify the location of these pools, and even more so for their individual participants, we do know that the cryptocurrency exchanges Binance and Huobi were both initially based in China but moved their headquarters abroad later on.

\begin{figure}[!ht]
\centering
{\includegraphics[width=\linewidth]{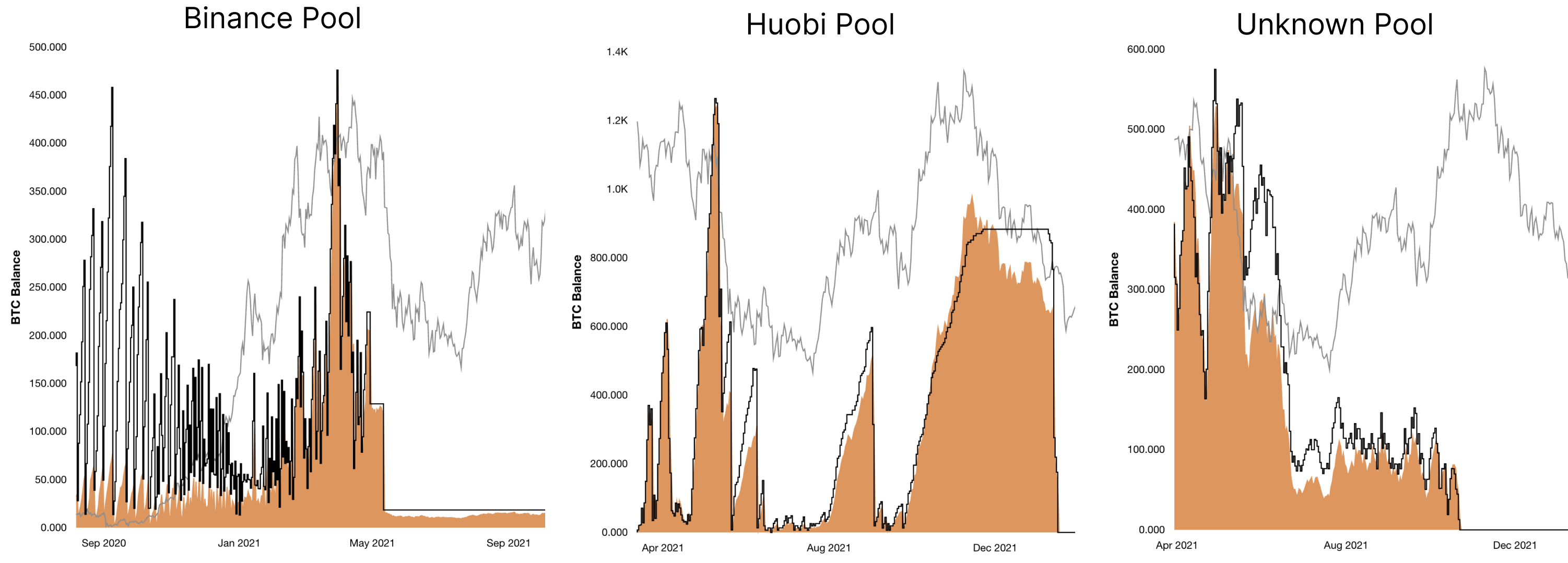}}
\caption{\small The BTC balance of major Bitcoin mining pools during China's mining ban.}
\label{fig_china_pools}
\end{figure}

Figure~\ref{fig_china_pools} displays the balance of BTC in the pools, obtained from \href{https://www.blockchain.com/}{blockchain.com}, as a proxy for mining activity, since mining results in an increased balance of BTC. We observe a clear, lasting reduction in the BTC balance in May 2021 for the Binance pool. For the Huobi Pool, we saw a slight reduction in BTC balance in May 2021 and a near-complete reduction from June to the end of July, which directly matches our analysis of total hashrates and shock exposure in Section~\ref{sec_variable}. For the Unknown Pool, we see a similar but less pronounced decrease in the BTC balance in May 2021 and then a larger decrease in July 2021. In contrast to these mining pools, some mining pools, such as those indicated by the prominent, rising blue and red lines in Figure~\ref{fig_china_mining}A, see even greater mining activity after the ban. The observed variations in the trends of mining pools not only align with our analysis of the impact of resource flexibility on decentralization in response to a major policy shock, but they also reveal the complex dynamics of blockchain decentralization in response to policy shocks. 

As an interesting side note, the mining pools that experienced greater activity after the ban are labeled as those of \href{https://www.blockchain.com/explorer/addresses/btc/12dRugNcdxK39288NjcDV4GX7rMsKCGn6B}{Antpool} and \href{https://www.blockchain.com/explorer/addresses/btc/18cBEMRxXHqzWWCxZNtU91F5sbUNKhL5PX}{ViaBTC}. Both companies were, and still are, headquartered in China. In contrast, Binance and Huobi, both of which experienced marked decreases in mining (Figure~\ref{fig_china_pools}), are no longer based in China, which suggests a consolidation of Bitcoin mining among Chinese miners. Whether or not such consolidation of Bitcoin mining was deliberate or happenstance, it may further support our resource flexibility hypothesis in either case. A deliberate consolidation would require high frictions, such as the costs and logistical challenges of relocating specialized ASIC hardware and significant electricity consumption, to prevent miners from migrating out of the jurisdiction. On the other hand, if the consolidation was happenstance, it still underscores the rigidity and lack of adaptability in less flexible systems like Bitcoin. Either way, the observed consolidation among Chinese miners lends further credence to our resource flexibility hypothesis, highlighting the role of resource flexibility in shaping the decentralization dynamics of blockchain networks.

\setcounter{figure}{0} 
\setcounter{table}{0} 
\setcounter{equation}{0} 
\clearpage

\subsection{Lagged Effects of China's mining ban} \label{sec_appendix_lagged_china}

Table~\ref{tb_china_lags} shows the coefficients of the lagged difference-in-difference estimation for China's mining ban according to Equation~\ref{eq_did_shift}.

\begin{table}[ht]
\centering
\footnotesize
\begin{threeparttable}
\begin{tabular}{@{\extracolsep{10pt}}lcccccc}
\hline
\hline \\[-1.8ex]
\textit{Dependent variable}     & \textit{Entropy} & \textit{Nodes} & \textit{Gini} & \textit{Nakamoto} & \textit{HHI} \\
\hline \\[-1.8ex]
\textbf{After}  & 0.212$^{***}$ & 12.114 & -0.039$^{***}$ & 0.332$^{**}$ & -0.018$^{**}$\\ 
 & (0.054) & (16.931) & (0.007) & (0.126) & (0.005)\\
\textbf{Bitcoin}  & 0.104$^{*}$ & -32.609$^{***}$ & -0.329$^{***}$ & 1.765$^{***}$ & -0.049$^{***}$\\ 
 & (0.043) & (1.077) & (0.010) & (0.140) & (0.005)\\
\textbf{Intercept}  & 3.609$^{***}$ & 53.090$^{***}$ & 0.809$^{***}$ & 2.660$^{***}$ & 0.146$^{***}$\\ 
 & (0.038) & (0.851) & (0.005) & (0.112) & (0.005)\\
\textbf{Treated$_{\tau=0}$}  & -0.224$^{***}$ & -12.614 & 0.027$^{*}$ & -0.299 & 0.018$^{**}$\\ 
 & (0.065) & (16.938) & (0.010) & (0.174) & (0.006)\\
\textbf{Treated$_{\tau=120}$}  & -0.140$^{***}$ & -1.267$^{**}$ & 0.015$^{***}$ & -0.533$^{***}$ & 0.012$^{***}$\\ 
 & (0.033) & (0.471) & (0.004) & (0.110) & (0.002)\\
\textbf{Treated$_{\tau=180}$}  & -0.360$^{***}$ & -4.251$^{***}$ & -0.020$^{***}$ & -0.438$^{***}$ & 0.023$^{***}$\\ 
 & (0.054) & (0.684) & (0.006) & (0.118) & (0.004)\\
\textbf{Treated$_{\tau=60}$}  & -0.041 & -1.067$^{**}$ & -0.009 & 0.017 & 0.001\\ 
 & (0.027) & (0.368) & (0.008) & (0.117) & (0.002)\\
\textbf{Treated$_{\tau=-120}$}  & 0.082$^{***}$ & 1.017$^{***}$ & -0.001 & 0.017 & -0.006$^{**}$\\ 
 & (0.022) & (0.234) & (0.003) & (0.124) & (0.002)\\
\textbf{Treated$_{\tau=-180}$}  & -0.008 & -0.998 & -0.000 & 0.208 & -0.000\\ 
 & (0.025) & (0.677) & (0.009) & (0.149) & (0.002)\\
\textbf{Treated$_{\tau=-60}$}  & -0.023 & 0.600 & 0.026$^{*}$ & -0.233 & 0.004\\ 
 & (0.034) & (0.464) & (0.012) & (0.172) & (0.003)\\
\hline \\[-1.8ex]
Observations                        & 2002 \\
\hline
\hline
\end{tabular}
\begin{tablenotes}
\small
\item Notes: $^{*}$p$<$0.05; $^{**}$p$<$0.01; $^{***}$p$<$0.001. Standard errors are clustered by blockchain and month.
\end{tablenotes}
\end{threeparttable}
\vspace{1em}
\caption{Lagged effects of China's ban on crypto mining.} \label{tb_china_lags}
\end{table}

\begin{figure}[!ht]
\centering
{\includegraphics[width=0.5\linewidth]{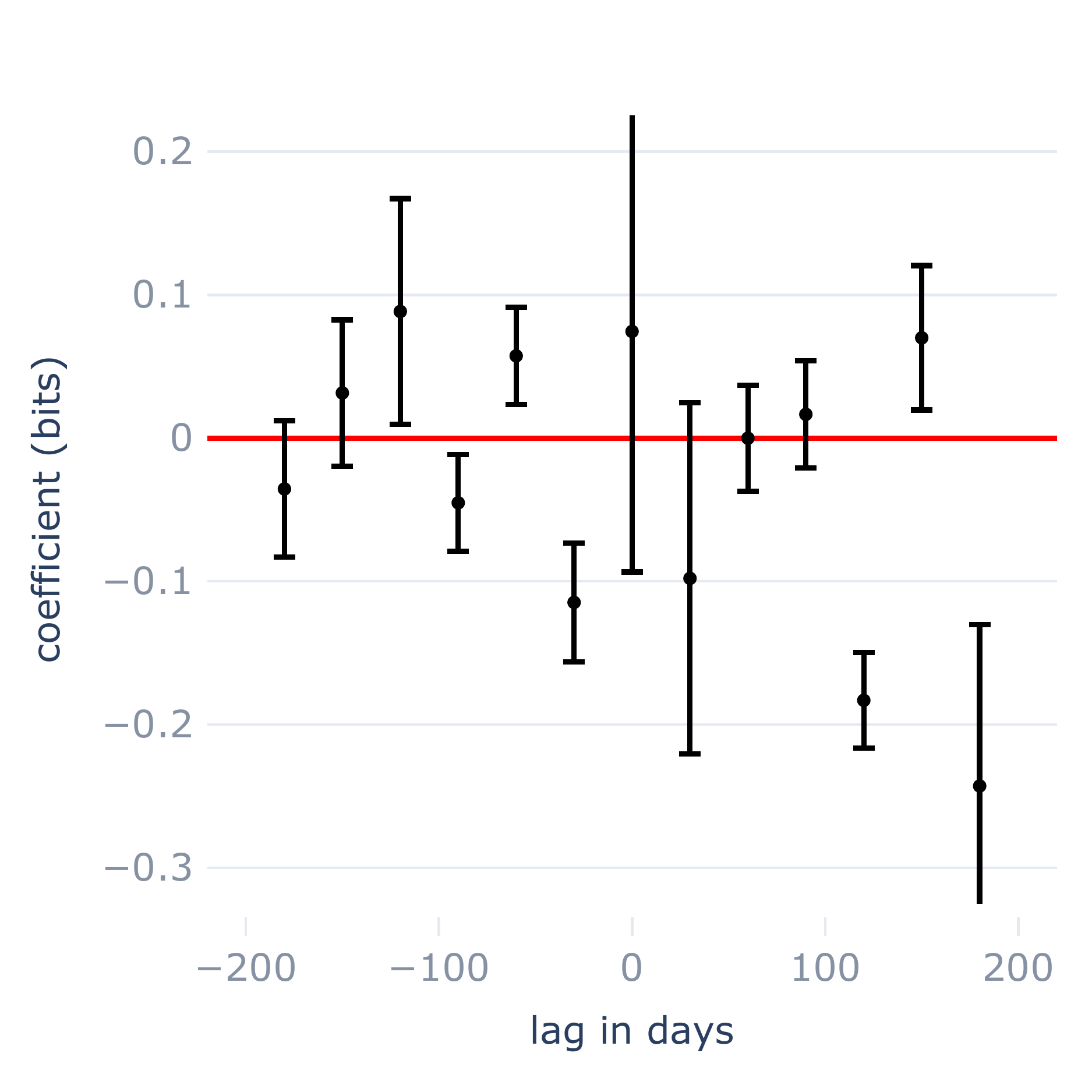}}
\caption{\small Coefficients for the lead and lag effects $\text{Treatment}_\lambda$ on the entropy of China's mining ban with 50-day intervals. Error bars indicate 95\% confidence intervals.} 
\label{fig_lag_china_50}
\end{figure}

We also conduct lagged analyses with a finer timescale of 50 days in Figure~\ref{fig_lag_china_50}.

\setcounter{figure}{0} 
\setcounter{table}{0} 
\setcounter{equation}{0} 
\clearpage



\subsection{Lagged Effects of Hetzner's Shutdown} \label{sec_appendix_lagged_hetzner}

Table~\ref{tb_hetzner_lags} shows the coefficients of the lagged difference-in-difference estimation for Hetzner's shutdown of Solana nodes according to Equation~\ref{eq_did_shift}.

\begin{table}[ht]
\centering
\begin{threeparttable}
\begin{tabular}{lccccc}
\hline
\hline \\[-1.8ex]
\textit{Dependent variable} & \textit{Entropy} & \textit{Nodes} & \textit{Gini} & \textit{Nakamoto} & \textit{HHI} \\
\hline \\[-1.8ex]
\textbf{Treated$_{\tau=0}$}  & -0.392$^{***}$ & -343.680$^{***}$ & 0.039 & -6.909$^{***}$ & 0.005\\ 
 & (0.113) & (61.343) & (0.052) & (1.775) & (0.004)\\
\textbf{Treated$_{\tau=10}$}  & 0.194$^{***}$ & 75.900 & -0.037$^{***}$ & -0.300 & -0.000$^{**}$\\ 
 & (0.038) & (42.539) & (0.005) & (0.630) & (0.000)\\
\textbf{Treated$_{\tau=20}$}  & 0.148$^{***}$ & 5.600 & -0.023$^{***}$ & 6.000$^{***}$ & -0.000$^{***}$\\ 
 & (0.020) & (5.144) & (0.003) & (0.887) & (0.000)\\
\textbf{Treated$_{\tau=30}$}  & 0.037 & 11.527$^{*}$ & 0.001 & 2.791$^{**}$ & -0.000$^{***}$\\ 
 & (0.021) & (5.240) & (0.004) & (0.928) & (0.000)\\
\textbf{Treated$_{\tau=-10}$}  & 0.011 & 25.300$^{***}$ & 0.000 & 0.400 & -0.000$^{*}$\\ 
 & (0.007) & (4.739) & (0.001) & (0.308) & (0.000)\\
\textbf{Treated$_{\tau=-20}$}  & 0.015$^{*}$ & 31.300$^{***}$ & -0.002 & 0.400 & -0.000\\ 
 & (0.007) & (5.892) & (0.001) & (0.287) & (0.000)\\
\textbf{Treated$_{\tau=-30}$}  & 0.009 & 53.400$^{***}$ & 0.002 & 0.300 & 0.000\\ 
 & (0.013) & (7.932) & (0.002) & (0.341) & (0.000)\\
\hline \\[-1.8ex]
Observations & 162 & 162 & 162 & 162 & 162 \\
\hline
\hline
\end{tabular}
\begin{tablenotes}
\small
\item Notes: $^{*}$p$<$0.05; $^{**}$p$<$0.01; $^{***}$p$<$0.001. Standard errors are clustered by blockchain and day.
\end{tablenotes}
\end{threeparttable}
\vspace{1em}
\caption{Lagged effects of Hetzner's shutdown of Solana nodes.} \label{tb_hetzner_lags}
\end{table}

\setcounter{figure}{0} 
\setcounter{table}{0} 
\setcounter{equation}{0} 
\clearpage

\subsection{Synthetic Difference-in-Difference for the Hetzner Shutdown with Multiple Bandwidths} \label{sec_appendix_hetzner_bandwidth}

Table~\ref{tb_hetzner_bandwidths} shows the coefficients of the synthetic difference-in-difference estimation for Hetzner's shutdown of Solana nodes according to Equation~\ref{eq_sdid} using multiple bandwidths in days before and after the shock on November 2, 2022.

\begin{table}[ht]
\centering
\begin{threeparttable}
\begin{tabular}{lccccc}
\hline \hline \\[-1.8ex]
\textit{Dependent variable} & $\pm$10 days      & $\pm$20 days  & $\pm$30 days  & $\pm$40 days & $\pm$50 days \\
    & (1) & (2) & (3) & (4) & (5) \\
\hline
\rowcolor{LightGray} Panel A: Entropy & & & & & \\ \hline \\[-2.2ex]
\textbf{Treatment} & -0.363 & -0.271 & -0.180 & -0.142 & -0.156\\
 & (0.734) & (0.727) & (0.696) & (0.557) & (0.505)\\
\hline
\rowcolor{LightGray} Panel B: Nodes & & & & & \\ \hline \\[-2.2ex]
\textbf{Treatment} & -332.145 & -289.867 & -262.205 & -232.356 & -222.801\\
 & (302.324) & (298.490) & (286.137) & (229.482) & (207.284)\\
\hline
\rowcolor{LightGray} Panel C: Gini & & & & & \\ \hline \\[-2.2ex]
\textbf{Treatment} & 0.040 & 0.025 & 0.009 & -0.001 & -0.008\\
 & (0.352) & (0.348) & (0.333) & (0.266) & (0.243)\\
\hline
\rowcolor{LightGray} Panel D: Nakamoto & & & & & \\ \hline \\[-2.2ex]
\textbf{Treatment} & -6.433 & -6.632 & -4.456 & -2.739 & -1.977\\
 & (11.403) & (11.284) & (10.838) & (8.874) & (8.235)\\
\hline
\rowcolor{LightGray} Panel E: HHI & & & & & \\ \hline \\[-2.2ex]
\textbf{Treatment} & 0.002 & 0.002 & 0.003 & 0.005 & 0.006\\
 & (0.025) & (0.024) & (0.024) & (0.019) & (0.018)\\
\hline
\\[-1.8ex]
Observations & 84 & 164 & 244 & 323 & 403 \\
\hline
\hline
\end{tabular}
\begin{tablenotes}
\small
\item Notes: $^{*}$p$<$0.05; $^{**}$p$<$0.01; $^{***}$p$<$0.001. Standard errors are derived from placebo tests.
\end{tablenotes}
\end{threeparttable}
\vspace{1em}
\caption{Synthetic difference-in-difference estimation of Hetzner shutdown with multiple bandwidths.} \label{tb_hetzner_bandwidths}
\end{table}

\setcounter{figure}{0} 
\setcounter{table}{0} 
\setcounter{equation}{0} 
\clearpage

\subsection{Price Correlation between BTC and ETH} \label{sec_appendix_price_corr}

\begin{figure}[!ht]
\centering
{\includegraphics[width=\linewidth]{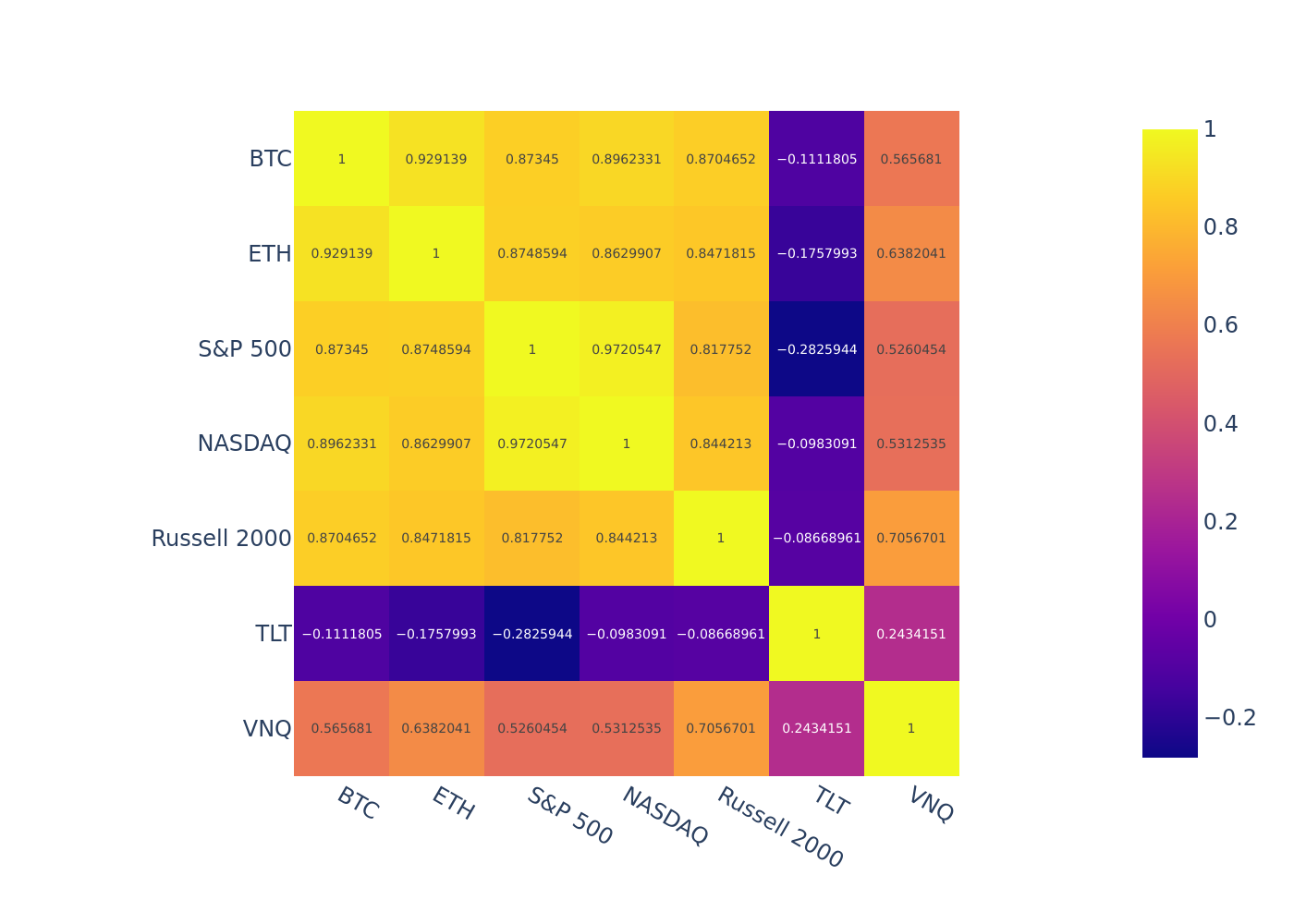}}
\caption{\small Pearson's correlation coefficients between BTC, ETH, the S\&P 500 Index, NASDAQ Composite, Russell 2000, TLT (the iShares 20+ Year Treasury Bond ETF), and VNQ (the Vanguard Real Estate Index Fund ETF).}
\label{fig_price_corr}
\end{figure}

Here, we measure the price correlation between BTC, ETH, and other financial assets to understand the level of price correlation between BTC and ETH. Using price data from Yahoo Finance, we find that BTC and ETH had a relatively high Pearson's correlation coefficient of 0.93 (Figure~\ref{fig_price_corr}).\footnote{Data obtained from Yahoo Finance for \href{https://finance.yahoo.com/quote/BTC-USD/history/}{BTC}, \href{https://finance.yahoo.com/quote/ETH-USD/history}{ETH}, \href{https://finance.yahoo.com/quote/\%5ESPX/history}{S\&P Index}, \href{https://finance.yahoo.com/quote/\%5EIXIC/history}{NASDAQ Composite}, \href{https://finance.yahoo.com/quote/\%5ERUT/history}{Russell 2000}, \href{https://finance.yahoo.com/quote/TLT/history}{TLT}, and \href{https://finance.yahoo.com/quote/VNQ/history}{VNQ}. Accessed April 5, 2024.} This is in comparison with correlations of 0.89 and 0.86 with the S\&P 500 for BTC and ETH, respectively. Correlations are shown for other prominent financial assets in Figure~\ref{fig_price_corr} for comparison. TLT is the iShares 20+ Year Treasury Bond ETF, and VNQ is the Vanguard Real Estate Index Fund ETF.

\setcounter{figure}{0} 
\setcounter{table}{0} 
\setcounter{equation}{0} 
\clearpage

\subsection{Price Analysis during Shocks} \label{sec_appendix_price}

Here, we test the hypothesis that market sentiment influences a blockchain's recovery from shocks. Below are estimations of Equation~\ref{eq_event} on the daily price of tokens native to the blockchains studied here. Using price data around key events for BTC, SOL, and ETH, we find that these shocks had no positive impact on prices; some even dropped significantly.
\begin{table}[ht]
\centering
\begin{threeparttable}
\begin{tabular}{@{\extracolsep{5pt}}lcccc} 
\hline \hline \\[-1.8ex] 
\textit{Dependent variable}     & \multicolumn{3}{c}{\textit{Price}} \\
                                & BTC               & SOL               & ETH \\
\hline \\[-1.8ex]
\textbf{Intercept} & 55929.964$^{***}$ & 30.415$^{***}$ & 1674.996$^{***}$\\
 & (770.736) & (0.488) & (28.007)\\
\textbf{After} & -14880.065$^{***}$ & -7.246$^{***}$ & -351.507$^{***}$\\
 & (1600.222) & (1.928) & (36.226)\\
\textbf{Days} & -25.735 & -0.054$^{***}$ & 0.713\\
 & (17.386) & (0.014) & (0.841)\\
\textbf{After $\times$ Days} & -121.117$^{**}$ & -0.201$^{***}$ & 0.936\\
 & (38.166) & (0.046) & (1.314)\\
\hline \\[-1.8ex]
Observations & 121 & 121 & 121 \\
\hline
\hline 
\end{tabular}
\begin{tablenotes}
\small
\item Notes: $^{*}$p$<$0.05; $^{**}$p$<$0.01; $^{***}$p$<$0.001. Heteroskedasticity-robust standard errors. Each column is a separate single-blockchain regression.
\end{tablenotes}
\end{threeparttable}
\vspace{1em}
\caption{Event studies on the token price of BTC during China's ban, SOL during Hetzner's shutdown, and ETH during the Merge.} \label{tb_price}
\end{table}

\setcounter{figure}{0} 
\setcounter{table}{0} 
\setcounter{equation}{0} 
\clearpage

\subsection{Clustering at the blockchain level} \label{sec_appendix_clustering}

As a robustness check, we demonstrate that clustering only at the blockchain level, not the blockchain-month level, results in unrealistically low standard errors, which would lead to overconfident statistical significance \citep{cameron2008bootstrap, petersen2009estimating}. Table~\ref{tb_china_blockchain_clustering} reproduces the results of Table~\ref{tb_china} with clustering at the blockchain level.

\begin{table}[ht]
\centering
\resizebox{\textwidth}{!}{
\begin{threeparttable}
\begin{tabular}{lcccccccccc}
\hline \\[-1.8ex]
\textit{Dependent} & \multicolumn{2}{c}{\textit{Entropy}} & \multicolumn{2}{c}{\textit{Nodes}} & \multicolumn{2}{c}{\textit{Gini}} & \multicolumn{2}{c}{\textit{Nakamoto}} & \multicolumn{2}{c}{\textit{HHI}}\\
\textit{variable} & (1) & (2) & (3) & (4) & (5) & (6) & (7) & (8) & (9) & (10) \\
\hline \\[-1.8ex]
\textbf{After}  & 0.064 & 0.072 & -3.093 & -3.203 & -0.033 & -0.034 & 0.241 & 0.251 & -0.004 & -0.004\\ 
 & (0.126) & (0.119) & (3.430) & (3.229) & (0.045) & (0.042) & (0.243) & (0.229) & (0.008) & (0.007)\\
\textbf{Exposure}  &  & -0.032 &  & 0.424 &  & 0.003 &  & -0.038 &  & 0.002\\ 
 &  & (0.029) &  & (0.784) &  & (0.010) &  & (0.056) &  & (0.002)\\
\textbf{Bitcoin}  & 0.020$^{***}$ & 0.020$^{***}$ & -33.980$^{***}$ & -33.980$^{***}$ & -0.327$^{***}$ & -0.327$^{***}$ & 1.577$^{***}$ & 1.577$^{***}$ & -0.037$^{***}$ & -0.037$^{***}$\\ 
 & (0.000) & (0.000) & (0.000) & (0.000) & (0.000) & (0.000) & (0.000) & (0.000) & (0.000) & (0.000)\\
\textbf{Treatment}  & -0.209$^{***}$ & -0.201$^{***}$ & 5.050$^{***}$ & 4.943$^{***}$ & 0.047$^{***}$ & 0.047$^{***}$ & -0.414$^{***}$ & -0.404$^{***}$ & 0.012$^{***}$ & 0.012$^{***}$\\ 
 & (0.000) & (0.007) & (0.000) & (0.198) & (0.000) & (0.003) & (0.000) & (0.014) & (0.000) & (0.000)\\
\textbf{Intercept}  & 3.672$^{***}$ & 3.672$^{***}$ & 52.990$^{***}$ & 52.990$^{***}$ & 0.791$^{***}$ & 0.791$^{***}$ & 3.069$^{***}$ & 3.069$^{***}$ & 0.138$^{***}$ & 0.138$^{***}$\\ 
 & (0.090) & (0.090) & (0.351) & (0.351) & (0.018) & (0.018) & (0.286) & (0.286) & (0.013) & (0.013)\\
\hline \\[-1.8ex]
Observations                & 1202 & 1202 & 1202 & 1202 & 1202 & 1202 & 1202 & 1202 & 1202 & 1202 \\
\hline
\hline
\end{tabular}
\begin{tablenotes}
\small
\item Notes: $^{*}$p$<$0.05; $^{**}$p$<$0.01; $^{***}$p$<$0.001. Standard errors are clustered by blockchain.
\end{tablenotes}
\end{threeparttable}
}
\vspace{1em}
\caption{China bans crypto mining} \label{tb_china_blockchain_clustering}
\end{table}

\setcounter{figure}{0}
\setcounter{table}{0}
\setcounter{equation}{0}
\clearpage

\subsection{Hetzner Shutdown Details} \label{sec_appendix_hetzner_details}

The reasons behind Hetzner's selective enforcement against Solana nodes, as opposed to nodes from other blockchains, remain unexplained by the company. Hetzner's public communications state a broad prohibition against the use of their services for any crypto-related activities, encompassing both PoW and PoS mechanisms, as well as indirectly related activities such as cryptocurrency trading. The absence of similar actions against other chains during the same period suggests a potentially unique issue with Solana's relatively high resource usage that triggered enforcement. For example, Solana nodes require 256GB of RAM, while Ethereum nodes require just 8 GB of RAM.\footnote{See hardware requirements for \href{https://docs.solanalabs.com/operations/requirements}{Solana} and \href{https://ethereum.org/en/run-a-node/}{Ethereum}. Accessed April 14, 2024.} If indeed Solana was targeted due to its excessive resource usage, this would support the role of resource efficiency of consensus mechanisms in enabling sustained blockchain decentralization.

\begin{figure}[!ht]
\centering
{\includegraphics[width=0.8\linewidth]{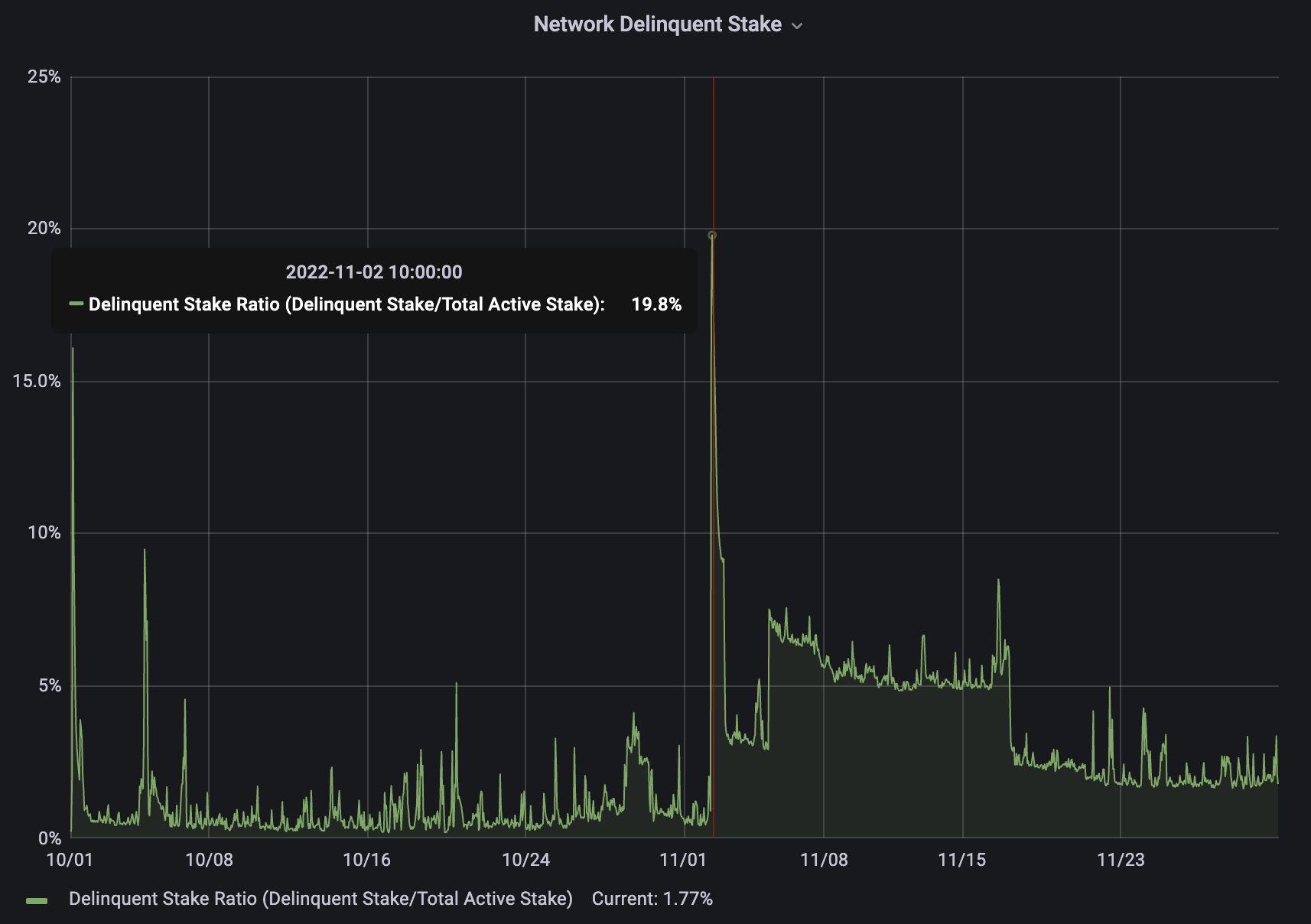}}
\caption{\small Delinquent stake ratio for Solana during the Hetzner shutdown.}
\label{fig_sol_delinquency}
\end{figure}

\setcounter{figure}{0}
\setcounter{table}{0}
\setcounter{equation}{0}
\clearpage

\subsection{Ethereum Merge Exogeneity} \label{sec_appendix_merge_exogeneity}

We have several reasons to believe the exogeneity of the Merge as a technical shock. First, the Merge was a literal merging of Ethereum's execution layer with a new, external Proof-of-Stake consensus layer, called the Beacon chain, which has been live and rewarding validators since December 1, 2020, but without validating any transactions on Ethereum.\footnote{See more on the Beacon chain on \href{https://ethereum.org/en/roadmap/beacon-chain/}{ethereum.org}. Accessed April 22, 2024.} The Merge simply replaced the Proof-of-Work consensus layer with the existing PoS Beacon chain. The Merge occurred literally in the span of less than 20 seconds and did not require changes to the network before or after the Merge.\footnote{See block 15,537,394 on \href{https://etherscan.io/block/15537394}{etherscan.io}. Accessed April 22, 2024.} Second, while the Merge had been announced in advance, it has been delayed five times since 2017 across five forks: Byzantium in 2017, Muir Glacier in 2020, London in August 2021, Arrow Glacier in December 2021, and finally Gray Glacier in June 2022.\footnote{See the forks of Ethereum on \href{https://ethereum.org/en/history/}{ethereum.org}. Accessed April 22, 2024.} Third, the Merge was a technical upgrade of nearly unprecedented scale or complexity in the blockchain ecosystem. Together, these factors highlight the difficulty of anticipating any effect on the Ethereum ecosystem on September 15, 2022, by participants.

\setcounter{figure}{0}
\setcounter{table}{0}
\setcounter{equation}{0}
\clearpage

\subsection{Data Collection Details} \label{sec_appendix_data_details}

The hashrate is a measure of the collective computing power in a Proof-of-Work blockchain network; specifically, it is the total number of hash functions that all nodes in a network compute per second. While it cannot be directly measured, it is estimated from the number of blocks currently being mined and the current block difficulty. The computational power of an individual node can be affected by gradual hardware degradation, hardware upgrades, changes in efficiency due to electricity or cooling, or, most importantly for our context, changes in the amount of resources used by the node, which can be hash power or blockchain tokens. Most crypto miners are profit-maximizing and thus operate at maximum efficiency given the high costs of ASICs and GPUs, even in geographic regions with cheap electricity.\footnote{See \href{https://compassmining.io/education/bitcoin-miners-are-moving-to-different-continents-in-search-of-cheaper-energy/}{compassmining.io}.}

\setcounter{figure}{0}
\setcounter{table}{0}
\setcounter{equation}{0}
\clearpage

\subsection{Additional Decentralization Measures} \label{sec_appendix_measures}

The Gini coefficient $\text{Gini}_{it}$ is a measure of inequality within a distribution and quantifies the disparity in block production across nodes in blockchain $i$ on day $t$. Formally, it is defined as
\begin{equation*}
\text{Gini}_{it} = \frac{\sum_{k=1}^{N_{it}} \sum_{l=1}^{N_{it}} |x_k - x_l|}{2N_{it}^2 \overline{x}_{it}} \text{,}
\end{equation*}
where $x_k$ and $x_l$ are values from $X_{it}$ and $\overline{x}_{it}$ is the average of all $x_k$ in $X_{it}$. It is unitless and is bounded by $[0,1]$. A higher Gini coefficient means less decentralization.\footnote{The Gini coefficient can be misleading without also considering the number of nodes. For example, consider the case when there is only one node (\textit{e.g.}, a dictatorship is one-man-one-vote). The Gini coefficient would then be zero, indicating full decentralization, but it is merely ``fully equal'' for only the single node. Hence, we consider it important to use multiple metrics when possible and Shannon entropy when one requires a single metric to encapsulate both the number of entities and the distribution of power over those entities.}

The Nakamoto coefficient $\text{Nakamoto}_{it}$ measures the minimum number of nodes required to collectively control over 51\% of the block production on a given day. Formally, it is defined as
\begin{equation*}
\text{Nakamoto}_{it}=\min \{ n \in [1,..., N_{it}] : \sum_{k=1}^n p_k \ge 0.51 \} \text{,}
\end{equation*}
where $p_k$ is the fraction of blocks produced by the $k^\text{th}$ node on day $t$ \citep{srinivasan_lee_2017}. The Nakamoto coefficient is a key measure for assessing the resilience of the blockchain against potential collusion or majority attacks. It has units of nodes and is bounded by $[1,N]$. A higher Nakamoto coefficient means more decentralization.

Finally, $\text{HHI}_{it}$ indicates the concentration of block production among nodes in blockchain $i$ on day $t$. Formally, it is defined as
\begin{equation*}
\text{HHI}_{it}=\sum_{k \in K_{it}}p_k^2 \text{.}
\end{equation*}
$\text{HHI}_{it}$ is calculated as the sum of the squares of individual market shares of all participants. It is unitless and is bounded by $[1/N,1]$. A higher HHI means less decentralization.

\setcounter{figure}{0}
\setcounter{table}{0}
\setcounter{equation}{0}
\clearpage

\section*{Appendix References}

\end{document}